\documentclass[journal]{IEEEtran}
\usepackage{amssymb}
\usepackage{amsmath}
\usepackage{graphicx}
\usepackage{extarrows}

\begin{document}

\title{Supervisory Control of Fuzzy Discrete Event Systems for Simulation Equivalence}

\author{Weilin Deng
        and Daowen Qiu$^{\star}$ 
\thanks{Weilin Deng is  with the Department of Computer Science, Sun Yat-sen University,
Guangzhou, 510006, China, and with the Department of Computer Engineering, Guangdong Industry Technical College, Guangzhou, 510300, China. (e-mail:williamten@163.com)}
\thanks{Daowen Qiu (Corresponding author) is with the Department of Computer science, Sun Yat-sen
 University, Guangzhou, 510006, China, and with SQIG--Instituto de Telecomunica\c{c}\~{o}es, Departamento de Matem\'{a}tica, Instituto Superior T\'{e}cnico,
  Technical University of Lisbon, Av. Rovisco Pais 1049-001, Lisbon, Portugal (e-mail: issqdw@mail.sysu.edu.cn)}}

\maketitle

\begin{abstract}
 The supervisory control theory of fuzzy discrete event systems (FDESs) for fuzzy language equivalence has been developed.
 However, in a way, language equivalence has limited expressiveness.
  So if the given specification can not be expressed by language equivalence, then the control for language equivalence does not work. In this paper, we further establish the supervisory control theory of FDESs for fuzzy simulation equivalence whose expressiveness is stronger than that of fuzzy language equivalence.
      First, we formalize the notions of fuzzy simulation and fuzzy simulation equivalence between two FDESs. Then we present a method for deciding whether there is a fuzzy simulation or not. In addition, we also show several basic properties of fuzzy simulation relations.
      Afterwards, we put forward the notion of fuzzy simulation-based controllability,
    and particularly show that it serves as a necessary and sufficient condition for the existence of the fuzzy  supervisors of FDESs. Moreover,
      we study the ``range" control problem of FDESs.
      Some examples are given to illustrate the main results obtained. 
\end{abstract}


\IEEEpeerreviewmaketitle

\newtheorem{definition}{Definition}
\newtheorem{theorem}{Theorem}
\newtheorem{lemma}{Lemma}
\newtheorem{corollary}{Corollary}
\newtheorem{remark}{Remark}
\newtheorem{algorithm}{Algorithm}
\newtheorem{example}{Example}
\newtheorem{proposition}{Proposition}

\section{Introduction}

\IEEEPARstart{D}{iscrete} event systems (DESs) are discrete states and dynamic event-driven systems.
There are many real-world systems that can be modeled as DESs, such as computer networks, transportation systems,
automated manufacturing systems and communication networks, etc. Supervisory control problem of DESs was first launched
by Ramadge and Wonham \cite{first-supevisory}. Since then, there is a considerable amount of literature on this
issue (for example, \cite{desbook}, \cite{kumar-books} and their references).
In Ramadge and Wonham's framework, 
   the objective of the control is to ensure that the controlled system is language equivalent with the given desired
  specification. \par
  Language equivalence preserves the safety properties of linear temporal logic (LTL), which has been used in modeling checking (for instance, \cite{li} and °²¶¨).  However, language equivalence is not adequate for describing the behavior equivalence of some nondeterministic systems \cite{zhou-2},
  so several notions of behavior equivalence which are finer than language equivalence, such as failures, refusal-trace,
  ready-trace, simulation \cite{zhou-3} and bisimulation \cite{zhou-2}, \cite{Liu-1} have been proposed. Notably,
  Zhou and Kumar \cite{zhou-3} investigated the supervisory control problem of
   nondeterministic DESs for simulation equivalence, whose control objective is to ensure the simulation equivalence of the controlled systems and
   the given specification.\par
   It should be pointed out that the DESs framework can only process crisp states and crisp states transitions.
However, in real-world situation, there are a large number of problems with vagueness, impreciseness, and subjectivity.

In order to characterize those properties in DESs, Lin and Ying \cite{Feng-1}
first proposed the fuzzy discrete event systems (FDESs) by combining fuzzy set theory \cite{zadeh} with classical DESs. In \cite{Feng-1}, FDESs are modelled as fuzzy automata.
It is worth pointing out that fuzzy automata were first discussed by Santos \cite{santos-1}, Wee \cite{wee-1}, Lee and Zadeh \cite{lee-1}. Since then, a growing body of literature has investigated this topic (we can refer to \cite{fuzzy-automata} and its references). By the way, fuzzy automata taking membership in a complete residuated lattice were studied in \cite{qiu-fuzzy1}, \cite{qiu-fuzzy2}, \cite{qiu-fuzzy3}. M. \'{C}iri\'{c} and his cooperators also studied fuzzy automata taking membership in a complete residuated lattice in a different view \cite{mc-fuzzy}, \cite{mc-2}, \cite{bi_fuzzy_automata}.
\par
Qiu \cite{Qiu-1},
as well as Cao and Ying \cite{Cao-1}, respectively, developed the supervisory control theory of FDESs with full observation. Further, Qiu and Liu, Cao and Ying studied the supervisory control issue of FDESs
with partial observation in \cite{Qiu-3} and \cite{cao-2}, and investigated the decentralized control issue
of FDESs in \cite{Liu-3} and \cite{cao-2}, respectively. Recently, Du, Ying and Lin provided a theory of extended FDESs for handling ranges
of knowledge uncertainties and subjectivity \cite{knoledge}. Jayasiri established modular supervisory control and hierarchical supervisory control theory of FDESs in \cite{Jayasiri-1}, and  generalized  the decentralized control theory of FDESs in \cite{g-de}.  Moreover, FDESs have been applied to practical problem in many areas, such as decision making \cite{decision-making}, clinical treatment planning \cite{treatplaning}, HIV/AIDS treatment regimen selection \cite{treatment-2}, \cite{treatment-3}, traffic management \cite{traffic} and mobile robot navigation \cite{robot2}, \cite{robot-3}, \cite{Jayasiri-2}, \cite{robot}, etc. Notably, FDES-based method for mobile robot navigation was compared with several different methods (including DES-based method, arbitration method, unmodulated method, etc.) by Jayasiri \cite{Jayasiri-2} and Rajibul Huq \cite{robot}, respectively. The results in  \cite{Jayasiri-2} and \cite{robot} reveal that FDES-based method has a superior performance over its classical counterparts, especially in complex environment.
\par
Nevertheless, it is necessary to mention that the works in  \cite{Qiu-1}, \cite{Qiu-3}, \cite{Liu-3}, \cite{Cao-1}, \cite{cao-2}, \cite{g-de}, \cite{Jayasiri-2}  are all based on fuzzy language equivalence. That is, the objective of the control is to ensure that the controlled system is fuzzy language equivalent with the given specification. Such type of control is usually called fuzzy language-equivalence control. Similarly, the control which ensures the controlled system is fuzzy simulation equivalent with the given specification is called fuzzy simulation-equivalence control.\par
 It is known that language equivalence has limited expressiveness, and the expressiveness of simulation equivalence is stronger than that of language equivalence \cite{zhou-3}. That is, there exist some problems that can be expressed by simulation equivalence but not by language equivalence. An example of such properties is: All paths contain a state starting from which all future states satisfy a certain property \cite{zhou-3}. If the given specification is like such a property, then the fuzzy language-equivalence control \cite{Qiu-1}, \cite{Qiu-3}, \cite{Cao-1}, \cite{cao-2} does not work and the fuzzy simulation-equivalence control is required (Example 5 in  Section  \uppercase\expandafter{\romannumeral4} intends to  illustrate this case).
 \par

  As far as we are aware, up to now, there are still no studies on the fuzzy simulation-equivalence control problem of FDESs.
  The purpose of this paper is to develop these works \cite{zhou-3}, \cite{Qiu-1}, \cite{Cao-1} and establish the fuzzy simulation-equivalence control theory for FDESs. In the paper, we are mainly concerned with what specifications can be achieved by fuzzy simulation-equivalence control and what are the relations between fuzzy language-equivalence control and fuzzy simulation-equivalence control of FDESs.

 \par

    The main contributions of the paper are as follows.\par
\begin{enumerate}
  \item To characterize the fuzziness of the simulation relation of FDESs, in Section \uppercase\expandafter{\romannumeral3} we present the formal definition of the fuzzy simulation relations of FDESs, which is the generalized version of the simulation relations of DESs. The generalization makes it possible to characterize the relations between FDESs more precisely.
      Then we present a method for deciding whether there is a fuzzy simulation between the given FDESs.
      Further we investigate several basic properties of fuzzy simulation relations, which are the foundations for the study of the fuzzy simulation-equivalence control problem of FDESs.
  \item In Subsection A of Section \uppercase\expandafter{\romannumeral4}, we introduce the notion of fuzzy simulation-based controllability and show that it serves as a necessary and sufficient
      condition for the existence of fuzzy supervisors of FDESs. We also present an efficient algorithm for constructing a fuzzy supervisor whenever it exists. Moreover, we study the ``range" control problem of FDESs and present a necessary and sufficient condition for the existence of the ``range" supervisors for FDESs.
  \item In Subsection B of Section \uppercase\expandafter{\romannumeral4}, we discuss the relations between the fuzzy language-equivalence control, which have been discussed by Qiu \cite{Qiu-1}, and the fuzzy simulation-equivalence control of FDESs. We find that the fuzzy simulation-based controllability implies the corresponding fuzzy language-based controllability but the inverse does not hold. This result suggests that the fuzzy simulation-equivalence control is more precise than the fuzzy language-equivalence control.
\end{enumerate}

    \par
    Besides the above-mentioned Sections, Section \uppercase\expandafter{\romannumeral2} provides the formulation of  FDESs with parallel composition operation, which has been introduced by Qiu \cite{Qiu-1}.
 Section \uppercase\expandafter{\romannumeral6} summarizes the main results obtained and mentions several future research directions. \par

\section{Fuzzy Discrete Event Systems}
    In this section, we would briefly introduce the language and automaton models for DESs and FDESs. For the more details,
    we can refer to \cite{Feng-1}, \cite{Qiu-1}, \cite{zhou-1}, \cite{desbook}. \par
    A DES is usually modeled by a finite automaton in logical level. A finite automaton is a 5-tuple $G=\{X, \Sigma, \alpha, X_{0}, X_{m}\}$, where $X$ denotes a set of finite states, $\Sigma$ denotes a set of events, $\alpha:X\times\bar{\Sigma}\rightarrow 2^{X}$ is the state transition function, where $\bar{\Sigma}=\Sigma\cup\{\epsilon\}$ with
    $\epsilon$ being a label for ``silent" transitions, $X_{0}\subseteq X$ is the set of initial states, and $X_{m} \subseteq X$ is the set of marked (final) states. $\Sigma^{*}$ denotes the set of all finite length sequences over $\Sigma$, including zero length string $\epsilon$.
    The $\epsilon$-closure of $x \in X$, denoted as $\epsilon^{*}(x)$, is the set of states reached by the execution of a sequence of
$\epsilon$-transitions from the state $x$. By using the $\epsilon$-closure map, we can extend the definition of the state transition function  to $X\times\Sigma^{*}$ in the following inductive manner: $
    \forall x\in X, \alpha^{*}(x, \epsilon)=\epsilon^{*}(x)$, and $\forall  s\in \Sigma^{*}, \sigma \in \Sigma:\alpha^{*}(x, s\sigma)=\epsilon^{*}(\alpha(\alpha^{*}(x, s), \sigma))$.
     A subset of $\Sigma^{*}$ is called a \emph{language}.  A language $K$ is closed if $K=pr(K)$, where $pr(K)$ denotes the prefix closure of $K$. The languages generated and marked by $G$ are, respectively, defined as $L(G)=\{s\in \Sigma:\alpha^{*}(X_{0}, s)\neq \emptyset \} $ and $L_{m}(G)=\{s\in L(G):\alpha^{*}(X_{0}, s)\bigcap X_{m}\neq \emptyset \} $.
    \par
      We consider the following vectors and matrices representation for FDESs \cite{Feng-1}, \cite{Qiu-1}.
    \begin{definition}
        An FDES is modeled as a fuzzy automaton, which is a max-min system:
        \[
        \tilde{G}=\{\tilde{X}, \tilde{\Sigma}, \tilde{\alpha}, \tilde{x}_{0}, \tilde{x}_{m}\}.
        \]
        Here $\tilde{X}$ is a set of fuzzy states over a crisp state set $X$ with  $|X| = n$.
         A fuzzy state $\tilde{x} \in \tilde{X}$ is represented by a vector $[x_{1}, x_{2}, \ldots, x_{n}]$, where $x_{i}\in[0, 1]$  represents the degree of fuzzy state $\tilde{x}$ being crisp state $x_{i}$.
         $\tilde{x}_{0} = [\tilde{x}_{0,1},\tilde{x}_{0,2},\ldots,\tilde{x}_{0,n}]$ is the fuzzy initial state, where $\tilde{x}_{0,i} \in [0,1]$ is the degree of the crisp state $x_{i}$ belonging to initial states.
        $\tilde{x}_{m} = [\tilde{x}_{m,1},\tilde{x}_{m,2},\ldots,\tilde{x}_{m,n}]$ is the fuzzy final state, where $\tilde{x}_{m,i} \in [0,1]$ is the degree of the crisp state $x_{i}$ belonging to final states.
        $\tilde{\Sigma}$ is a set of fuzzy events. Any $\tilde{\sigma}\in \tilde{\Sigma}$ is denoted by a matrix $\tilde{\sigma}=[a_{ij}]_{n*n}$ with $a_{ij}\in [0, 1]$.
        $\alpha:\tilde{X}\times\tilde{\Sigma}\rightarrow \tilde{X}$ is a transition function, which is defined by $\tilde{\alpha}(\tilde{x},  \tilde{\sigma})=\tilde{x} \odot\tilde{\sigma}$ for $\tilde{x}\in \tilde{X}$ and $\tilde{\sigma} \in \tilde{\Sigma}$.
        The ``$\odot$" denotes max-min operation in fuzzy set theory \cite{fuzzybook}: For matrices $A=[a_{il}]_{n*m}$ and $B=[b_{lj}]_{m*k}$,  matrix $C=A\odot B=[c_{ij}]_{n*k}$ with $c_{ij}=\max_{l=1}^{m}\min\{a_{il}, b_{lj}\}$.
    \end{definition}
\begin{remark}
  The vectors and matrices representation for FDESs mentioned in Definition 1 is also suitable for  DESs. Actually, if we restrict all the elements in state vectors and event matrices to $\{0 , 1\}$, then an FDES is reduced to a  DES.
\end{remark}

    The fuzzy languages generated and marked by $\tilde{G}$, denoted by $L_{\tilde{G}}$ and $L_{\tilde{G}, m}$, respectively, are defined as two functions from $\tilde{\Sigma}$ to $[0, 1]$ as follows: $ L_{\tilde{G}}(\epsilon)= L_{\tilde{G},m}(\epsilon) = 1$, and for any fuzzy events string $\tilde{s}=\tilde{\sigma}_{1}\tilde{\sigma}_{2}\ldots\tilde{\sigma}_{k}\in \tilde{\Sigma}^{*}, k \geq 1$,
    \begin{equation}
      L_{\tilde{G}}(\tilde{s})=\max_{i=1}^{n}\tilde{x}_{0}\odot\tilde{\sigma}_{1}\odot\ldots\odot\tilde{\sigma}_{k}\odot \bar{s}_{i}^{T},
    \end{equation}
    \begin{equation}
      L_{\tilde{G}, m}(\tilde{s})=\tilde{x}_{0}\odot\tilde{\sigma}_{1}\odot\ldots\odot\tilde{\sigma}_{k}\odot \tilde{x}_{m}^{T},
    \end{equation}
    where T is the transpose operation, and $\bar{s}_{i}=[0\ldots1\ldots0]$, where 1 is at the $i$th entry. The following property is obtained in \cite{Qiu-1}. For any $\tilde{s}\in\tilde{\Sigma}^{*}$ and any $\tilde{\sigma}\in\tilde{\Sigma}$,
    \begin{equation}
      L_{\tilde{G}, m}(\tilde{s}\tilde{\sigma})\leq L_{\tilde{G}}(\tilde{s}\tilde{\sigma})\leq L_{\tilde{G}}(\tilde{s}).
    \end{equation}
    \begin{example}
    Let an FDES $\tilde{G}=\{\tilde{X}, \tilde{\Sigma}, \tilde{\alpha}, \tilde{x}_{0}, \tilde{x}_{m}\}$, $\tilde{\Sigma}=\{\tilde{\sigma}, \tilde{\sigma}^{'}\}$ and
   \[
   \tilde{x}_{0}=[0.9~0.1], \  \tilde{x}_{m}=[0.1~0.8],
   \]
   \[\tilde{\sigma}=\left[
                                                \begin{array}{ccc}
                                                   0.9   & 0. 8 \\
                                                   0  & 0.1 \\
                                                \end{array}
                                             \right], \
                                             \tilde{\sigma}^{'}=\left[
                                                \begin{array}{ccc}
                                                   0  & 0. 3 \\
                                                   0  & 0.9 \\
                                                \end{array}
                                             \right],
\]
where the $\tilde{\sigma}$ and $\tilde{\sigma}^{'}$ are the corresponding matrices of events $\tilde{\sigma}$ and $\tilde{\sigma}^{'}$, respectively.
$\tilde{G}$ is shown as Fig.1-(A). If we restrict all the elements in vectors $\tilde{x}_{0}$, $\tilde{x}_{m}$ and matrices $\tilde{\sigma}$, $\tilde{\sigma}^{'}$ to 0 or 1, for instance, we revise 0.9 and 0.8 to 1, 0.3 and 0.1 to 0, then the FDES $\tilde{G}$ is transformed into a nondeterministic DES $ G=\{X, \Sigma, \alpha, x_{0}, x_{m}\}$ (as shown in Fig.1-(B)), where
   \[
   x_{0}=[1~0], x_{m}=[0~1],\sigma=\left[
                                                \begin{array}{ccc}
                                                   1   & 1 \\
                                                   0  & 0 \\
                                                \end{array}
                                             \right],
                                             \sigma^{'}=\left[
                                                \begin{array}{ccc}
                                                   0  & 0 \\
                                                   0  & 1 \\
                                                \end{array}
                                             \right].
\]
According to Equations (1) and (2), the languages generated and marked by FDES $\tilde{G}$ are shown as follows.
\[
 L_{\tilde{G}} = \frac{1}{\epsilon} + \frac{0.9}{\tilde{\sigma}} +   \frac{0.3}{\tilde{\sigma}^{'}} +  \frac{0.9}{\tilde{\sigma}\tilde{\sigma}} +
\frac{0.8}{\tilde{\sigma}\tilde{\sigma}^{'}} + \frac{0.1}{\tilde{\sigma}^{'}\tilde{\sigma}} +\frac{0.3}{\tilde{\sigma}^{'}\tilde{\sigma}^{'}} \ldots,
\]
\[
 L_{\tilde{G},m} = \frac{1}{\epsilon} +  \frac{0.8}{\tilde{\sigma}} +   \frac{0.3}{\tilde{\sigma}^{'}} +  \frac{0.8}{\tilde{\sigma}\tilde{\sigma}} +
\frac{0.8}{\tilde{\sigma}\tilde{\sigma}^{'}} + \frac{0.1}{\tilde{\sigma}^{'}\tilde{\sigma}} +\frac{0.3}{\tilde{\sigma}^{'}\tilde{\sigma}^{'}} \ldots.
\]
The languages generated and marked by the DES $G$ are shown as follows.
\[
 L_{G} = \frac{1}{\epsilon} +  \frac{1}{{\sigma}} +   \frac{0}{{\sigma}^{'}} +  \frac{1}{{\sigma}{\sigma}} +
\frac{1}{{\sigma}{\sigma}^{'}} + \frac{0}{{\sigma}^{'}{\sigma}} +\frac{0}{{\sigma}^{'}{\sigma}^{'}} \cdots,\]
\[
 L_{G,m} = \frac{1}{\epsilon} +  \frac{1}{{\sigma}} +   \frac{0}{{\sigma}^{'}} +  \frac{1}{{\sigma}{\sigma}} +
\frac{1}{{\sigma}{\sigma}^{'}} + \frac{0}{{\sigma}^{'}{\sigma}} +\frac{0}{{\sigma}^{'}{\sigma}^{'}} \cdots.
\]
Therefore, actually, nondeterministic DESs are just special cases of FDESs.

\end{example}

\begin{figure}
\centering
\includegraphics[width=0.32\textwidth]{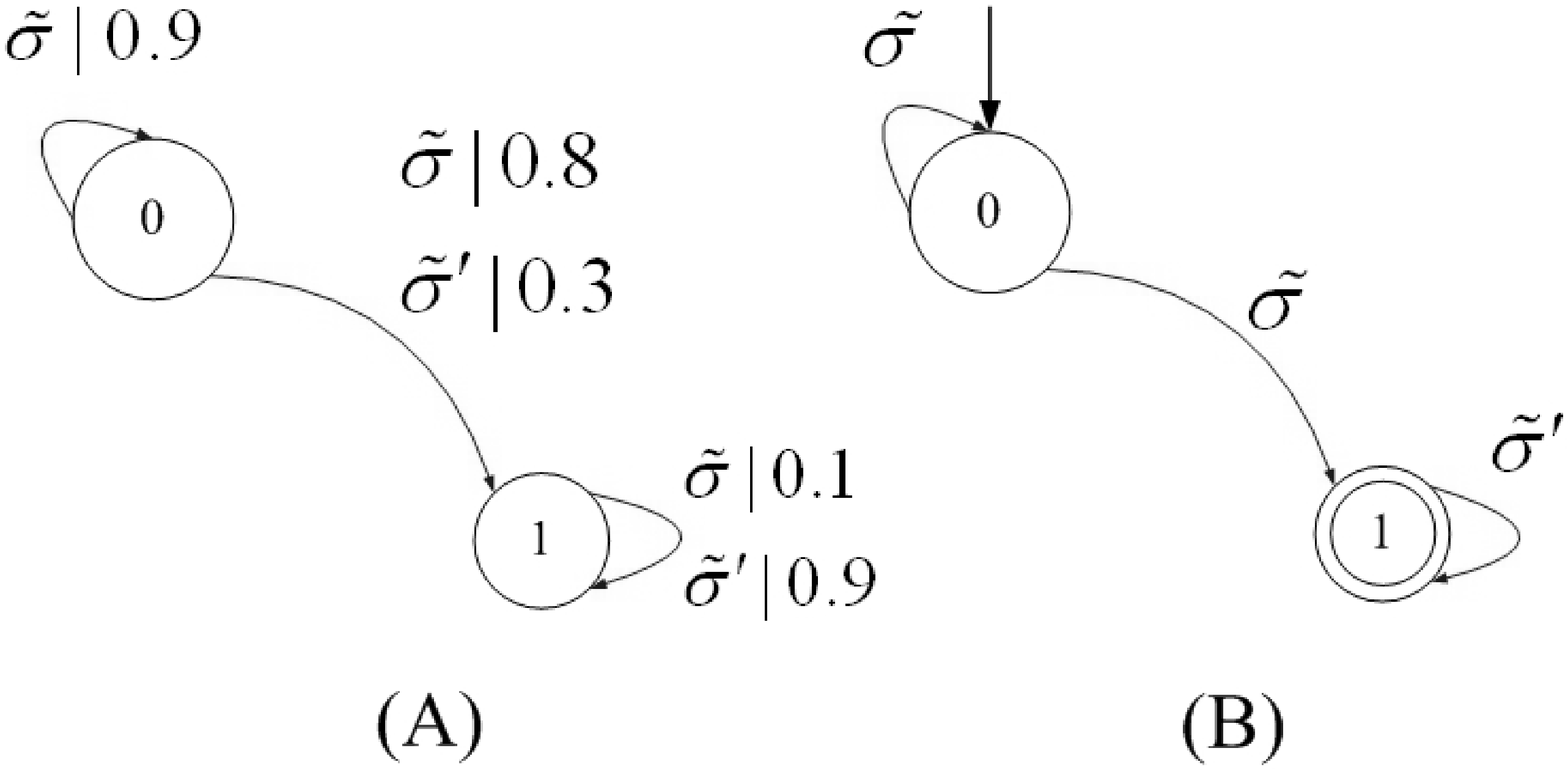}
\caption{\label{fig1} (A). FDES  $\tilde{G}$ of Example1.  (B). nondeterministic DES $G$ altered from $\tilde{G}$ by restricting the degree of transition and the degree of belonging to initial and final states to 0 or 1. }
\end{figure}
    \par
    In supervisory control theory, the operation of parallel composition is one of the most important operations over fuzzy automata. It characterizes the fuzzy systems combining with each other by synchronously executing the common events.
    For given $\tilde{G}_{i}=\{\tilde{X}_{i}, \tilde{\Sigma}_{i}, \tilde{\alpha}_{i}, \tilde{x}_{0i}, \tilde{x}_{mi}\}, i\in\{1, 2\}$, we formulate the parallel composition of fuzzy automata in terms of the following fashion:
    \[
    \tilde{G}_{1}||\tilde{G}_{2}=\{\tilde{X}_{1}\tilde{\otimes} \tilde{X}_{2}, \tilde{\Sigma}_{1}\cup\tilde{\Sigma}_{2}, \tilde{\alpha}_{1}||\tilde{\alpha}_{2}, \tilde{x}_{01}\tilde{\otimes} \tilde{x}_{02}, \tilde{x}_{1m}\tilde{\otimes} \tilde{x}_{2m} \}.
    \]
    Here $\tilde{X}_{1}\tilde{\otimes} \tilde{X}_{2}=\{\tilde{x}_{1}\tilde{\otimes} \tilde{x}_{2}: \tilde{x}_{i}\in \tilde{X}_{i}, i\in\{1, 2\} \}$, where $\tilde{\otimes}$ denotes fuzzy tensor operation. $\tilde{\alpha}_{1}||\tilde{\alpha}_{2}$ is a function from
    $(\tilde{X}_{1}\tilde{\otimes}\tilde{X}_{2})\times(\tilde{\Sigma}_{1}\cup\tilde{\Sigma}_{2})$ to $(\tilde{X}_{1}\tilde{\otimes}\tilde{X}_{2})$. That is, for any $\tilde{x}_{1}\tilde{\otimes} \tilde{x}_{2}\in(\tilde{X}_{1}\tilde{\otimes}\tilde{X}_{2})$ and any $\tilde{\sigma}\in(\tilde{\Sigma}_{1}\tilde{\otimes}\tilde{\Sigma}_{2})$,
    \[
        (\tilde{\alpha}_{1}||\tilde{\alpha}_{2})(\tilde{x}_{1}\tilde{\otimes} \tilde{x}_{2}, \tilde{\sigma})=(\tilde{x}_{1}\tilde{\otimes} \tilde{x}_{2})\odot\tilde{\sigma}.
    \]
    Here the corresponding matrix $\tilde{\sigma}$ of fuzzy event $\tilde{\sigma}$ is defined as follows.
    \begin{enumerate}
    \item If fuzzy event $\tilde{\sigma}\in\tilde{\Sigma}_{1}\cap\tilde{\Sigma}_{2}$, then the matrix $\tilde{\sigma} = \tilde{\sigma}_{1}\tilde{\otimes}\tilde{\sigma}_{2}$, where $\tilde{\sigma}_{1}$ and $\tilde{\sigma}_{2} $ are the corresponding matrices of fuzzy event $\tilde{\sigma}$ in $\tilde{G}_{1}$ and $\tilde{G}_{2}$, respectively.
    \item If fuzzy event $\tilde{\sigma}\in\tilde{\Sigma}_{1} \backslash \tilde{\Sigma}_{2}$, then the matrix $\tilde{\sigma} = \tilde{\sigma}_{1}\tilde{\otimes}\tilde{I}_{2}$, where $\tilde{\sigma}_{1}$ is the corresponding matrix of fuzzy event $\tilde{\sigma}$ in $\tilde{G}_{1}$, and $\tilde{I}_{2}$ is the unit matrix of order $|{X}_{2}|$.
    \item If fuzzy event $\tilde{\sigma}\in\tilde{\Sigma}_{2} \backslash \tilde{\Sigma}_{1}$, then the matrix $\tilde{\sigma} = \tilde{I}_{1}\tilde{\otimes}\tilde{\sigma}_{2}$, where $\tilde{\sigma}_{2}$ is the corresponding matrix of fuzzy event $\tilde{\sigma}$ in $\tilde{G}_{2}$, and $\tilde{I}_{1}$ is the unit matrix of order $|{X}_{1}|$.
    \end{enumerate}\par
    As indicated above, the symbol $\tilde{\otimes}$ denotes fuzzy tensor of matrices. That is, for matrices $A=[a_{ij}]_{m\times n}$ and  $B=[b_{ij}]_{k\times l}$, we have
    \[
    A\tilde{\otimes} B = \left[ \begin{array}{ccc}
            Min\{a_{11}, B\} & \ldots & Min\{a_{1n}, B\} \\
            \vdots & \ddots & \vdots\\
            Min\{a_{m1}, B\} & \ldots & Min\{a_{mn}, B\}
            \end{array} \right],
    \]
    where 
    \[
    Min(a_{ij}, B)=\left[ \begin{array}{ccc}
            \min\{a_{ij}, b_{11}\} & \ldots & \min\{a_{ij}, b_{1l}\} \\
            \vdots & \ddots & \vdots\\
            \min\{a_{ij}, b_{k1}\} & \ldots & \min\{a_{ij}, b_{kl}\}
            \end{array} \right].
    \]
    \begin{remark}
      The operation of parallel composition over fuzzy automata defined here is a little different from that in \cite{Qiu-1}.
      Namely, we use fuzzy tensor operation rather than tensor product operation.
      Such a choice can  ensure the correctness of the basic properties about the fuzzy simulation relations. We would discuss these properties in next section.
    \end{remark}

\section{Fuzzy Simulation and Fuzzy Simulation Equivalence of FDESs}
In this section, the notions of simulation and simulation equivalence of finite automata are generalized to their corresponding versions of fuzzy automata: fuzzy simulation and fuzzy simulation equivalence. Then we present a method for deciding whether
there is a fuzzy simulation between the given FDESs.  Furthermore, we discuss several basic properties of fuzzy simulation relations, which are the foundations for the study of the fuzzy simulation-equivalence control problem of FDESs.
\subsection{Fuzzy Simulation and Fuzzy Simulation Equivalence}
Firstly, we introduce the following two notions for finite automata, which have been presented in \cite{zhou-1,zhou-3}.
\begin{definition}
Given two finite automata $G_{i}=\{ X_{i}, \Sigma, \alpha_{i},$ $ X_{0i}, X_{mi}\}$, where $i\in\{1, 2\}$, $G_{1}$ is said to be \emph{simulated by} $G_{2}$, denoted by $G_{1}\subseteq_{\phi}G_{2}$, if there exists a binary relation $\phi\in \mathcal{P}(X_{1}\times X_{2})$ ($\mathcal{P}(\bullet)$ denotes the powerset of a set), which satisfies the following conditions:
\begin{subequations}
  \begin{gather}
  \begin{align}
     &(1).(\forall x_{01}\in X_{01})(\exists x_{02}\in X_{02}) ((x_{01}, x_{02})\in\phi),\\ 
     &(2).(\forall (x_{1}, x_{2})\in\phi)  \Rightarrow  (x_{1}\in X_{m1} \Rightarrow x_{2}\in X_{m2}),\\
     &(3).(\forall (x_{1}, x_{2})\in\phi) \Rightarrow (\forall \sigma\in\Sigma)(\forall x_{1}^{'}\in\alpha_{1}(x_{1}, \sigma)) \nonumber \\
     &(\exists x_{2}^{'}\in\alpha_{2}(x_{2}, \sigma))((x_{1}^{'}, x_{2}^{'})\in\phi).
     \end{align}
  \end{gather}
\end{subequations}
Here $\phi$ is called as a \emph{simulation relation}. For each $(x_{1}, x_{2})\in\phi$, $x_{1}$ is said to \emph{be simulated by} $x_{2}$. The subscript $\phi$ is omitted from $\subseteq_{\phi}$ when it is clear from the context.\par
Intuitively, the condition (1) shows that for each initial state of $G_{1}$, there exists at least one initial state of $G_{2}$ that can simulate the former; the condition (2) shows that marked states of $G_1$ can only be simulated by marked states of $G_{2}$; the condition (3) shows that for each simulation pair $(x_1,x_2) \in \phi$, each successor of $x_1$ can be simulated by at least one successor of $x_2$ under the same event-driven.
\end{definition}
\begin{definition}
Given two finite automata $G_{i}=\{ X_{i}, \Sigma, \alpha_{i},$ $ X_{0i}, X_{mi}\}$, where $i\in\{1, 2\}$, if there exist simulation relations $\phi_{1}\in \mathcal{P}(X_{1}\times X_{2})$ and $\phi_{2}\in \mathcal{P}(X_{2}\times X_{1})$ such that $G_{1}\subseteq_{\phi_{1}}G_{2}$ and $G_{2}\subseteq_{\phi_{2}}G_{1}$, $G_{1}$ is said to be \emph{simulation equivalent} with $G_{2}$, denoted by $G_{1}\sim_{\phi} G_{2}$, $\phi=\phi_{1}\cup\phi_{2}$. $\phi$ is called as a \emph{simulation equivalence relation}. The subscript $\phi$ is omitted from $\sim_{\phi}$ when it is clear from the context.
\end{definition}
\par

We consider to use a matrix to represent
  a simulation relation. Suppose $|{X}_{1}|=m, |X_{2}|=n$. Then the simulation relation $\phi$ can be represented by a matrix $[\phi_{ij}]_{m*n}$. The $\phi_{ij}\in\{0,1\}$, and $\phi_{ij} = 1$ if and only if $(x^{1}_{i},x^{2}_{j}) \in \phi$, where $x^{1}_{i}\in X_{1}$ and $x^{2}_{j}\in X_{2}$.
 The states that can simulate $x^{1}_{i}$ and the states that can be simulated by $x^{2}_{j}$ are denoted by $x^{1}_{i}\odot\phi$ and $x^{2}_{j}\odot\phi^{T}$, respectively. Here $x^{1}_{i}$ and $x^{2}_{j}$ are vectors as mentioned in Section 2.

 Then, we could use the vectors and matrices representation to reformulate the conditions in Definition 2 as follows:
\begin{enumerate}
\item
   Equation (4a) means that the initial states of $G_{1}$ should be included in the states that the initial states of $G_{2}$ can simulate. Hence, we reformulate the rule as: $X_{01}\leq (X_{02} \odot \phi^{T})$.
\item
 Equation (4b) means that the states that can simulate the final states of $G_{1}$ should be included in the final states of $G_{2}$.
 Hence, we reformulate the rule as: $(X_{m1} \odot \phi) \leq X_{m2}$.
\item
 Equation (4c) means that given any state $x_{2}\in X_{2}$, for any its simulation pair $(x_{1}, x_{2})$, suppose they make a same event-transition and turn to new states $X_{1*}$ and $X_{2*}$, respectively. Then $X_{1*}$ should be included in the states set that can be simulated by  $X_{2*}$. Hence,
we reformulate the rule as: $(x_{2}\odot \phi^{T} \odot \sigma_{1}) \leq (x_{2} \odot \sigma_{2} \odot \phi^{T})$ for any $ x_{2}\in X_{2}$ and any $ \sigma \in \Sigma$, where  $\sigma_{1}$ and $\sigma_{2}$ are the corresponding matrices of event $\sigma$ in $G_{1}$ and $G_{2}$, respectively.
Actually, the rule can be simplified as: $(\phi^{T} \odot \sigma_{1}) \leq (\sigma_{2} \odot \phi^{T})$ for any $\sigma \in \Sigma$.
\end{enumerate}

Similarly, the fuzzy simulation relations of FDESs also can be represented by a matrix. More precisely, consider fuzzy automata $\tilde{G}_{i}=\{\tilde{X}_{i}, \tilde{\Sigma}, \tilde{\alpha}_{i}, \tilde{x}_{0i}, \tilde{x}_{mi}\}$, where
$i\in\{1, 2\}$. Suppose $|{X}_{1}|=m, |X_{2}|=n$. Then the simulation relation between $\tilde{G}_{1}$ and $\tilde{G}_{2}$ is denoted by a fuzzy relation $\tilde{\phi}\in\mathcal{F}(X_{1}\times X_{2})$ ($\mathcal{F}(\bullet)$ denotes the set of all fuzzy subsets \cite{fuzzybook}). Let $\tilde{\phi}=[a_{ij}]_{m*n}, a_{ij}\in [0, 1]$, where $a_{ij}$ denotes the degree of the crisp state $x^{1}_{i}(\in X_{1})$ being simulated by the crisp state $x^{2}_{j}(\in X_{2})$. For any fuzzy state $\tilde{x}_{1}\in \tilde{X}_{1}$, the fuzzy state in $\tilde{G}_{2}$  that can simulate $\tilde{x}_{1}$ is denoted by $\tilde{x}_{1}\odot\tilde{\phi}$. On the other hand, for any state $\tilde{x}_{2}\in \tilde{X}_{2}$, the fuzzy state in $\tilde{G}_{1}$ that can be simulated by $\tilde{x}_{2}$ is denoted by $\tilde{x}_{2}\odot\tilde{\phi}^{T}$.
\par

Based on the above analysis, we present the formal definitions of \emph{fuzzy simulation} and \emph{fuzzy simulation equivalence}  as follows.
\par
\begin{definition}
    Given two fuzzy automata $\tilde{G}_{i} = \{\tilde{X}_{i}, \tilde{\Sigma}, \tilde{\alpha}_{i},$ $ \tilde{x}_{0i}, \tilde{x}_{mi} \}$, where
    $i\in\{1, 2\}$,
    $\tilde{G}_{1}$ is said to be \emph{fuzzy simulated by }$\tilde{G}_{2}$, denoted by $\tilde{G}_{1}\subseteq_{\tilde{\phi}}\tilde{G}_{2}$, if there exists a fuzzy relation $\tilde{\phi}\in\mathcal{F}(\tilde{X}_{1}\times \tilde{X}_{2})$, which satisfies the following conditions:
    \begin{subequations}\label{fuzzy simulation}
        \begin{gather}
            \begin{align}
              &(1). \tilde{x}_{01} \leq  \tilde{x}_{02}\odot \tilde{\phi}^{T}, \\
              &(2). \tilde{x}_{m1}\odot \tilde{\phi} \leq \tilde{x}_{m2}, \\
              &(3). \tilde{\phi}^{T} \odot \tilde{\sigma}_{1} \leq \tilde{\sigma}_{2}\odot \tilde{\phi}^{T} (\forall \tilde{\sigma} \in \tilde{\Sigma}).
            \end{align}
        \end{gather}
    \end{subequations}
    Here $\tilde{\sigma}_{1}$ and $\tilde{\sigma}_{2}$ denote the corresponding matrices of event $\tilde{\sigma}$ in $\tilde{G}_{1}$ and $\tilde{G}_{2}$, respectively.
      $\tilde{\phi}$ is called as a \emph{fuzzy simulation relation}.  Equations (5a)-(5c) are called as \emph{fuzzy simulation conditions} with respect to $\tilde{G}_1\rightarrow\tilde{G}_2$. The subscript $\tilde{\phi}$ is omitted from $\subseteq_{\tilde{\phi}}$ when it is clear from the context.
\end{definition}

\begin{definition}
Given two finite automata $\tilde{G}_{i}=\{ \tilde{X}_{i}, \tilde{\Sigma},$ $ \tilde{\alpha}_{i}, $ $\tilde{x}_{0i}, \tilde{x}_{mi}\}$, where $i\in\{1, 2\}$, if there exist fuzzy simulation relations $\tilde{\phi}_{1}\in \mathcal{F}(X_{1}\times X_{2})$ and $\tilde{\phi}_{2}\in \mathcal{F}(X_{2}\times X_{1})$ such that $\tilde{G}_{1}\subseteq_{\tilde{\phi}_{1}}\tilde{G}_{2}$ and
$\tilde{G}_{2}\subseteq_{\tilde{\phi}_{2}}\tilde{G}_{1}$, $\tilde{G}_{1}$ is said to be \emph{fuzzy simulation equivalent} with $\tilde{G}_{2}$, denoted by $\tilde{G}_{1}\sim \tilde{G}_{2}$.

\end{definition}
  \par
   By the way, Cao et al. \cite{cao-3} presented a notion related to the bisimulation for fuzzy-transition systems from a different point of view. Xing et al. \cite{xing} also defined simulation and bisimulation for fuzzy automata. However, the relation defined by Xing et al. \cite{xing} is actually crisp rather than fuzzy. Definition 4 generalizes the simulation relations by allowing the states of automata to simulate with any degree. This generalization makes it possible to characterize the relations between automata  more precisely. It should be pointed out that \emph{fuzzy simulation relations} defined here is equivalent with \emph{forward simulation relations} defined in \cite{bi_fuzzy_automata} for the particular max-min systems.\par
The following example illustrates the concepts defined before.
\begin{example}
   Let $\tilde{G}_{i}=\{\tilde{X}_{i}, \tilde{\Sigma}, \tilde{\alpha}_{i}, \tilde{x}_{0i}, \tilde{x}_{mi}\}, i\in\{1, 2\}$, where $\tilde{\Sigma}=\{\tilde{\sigma}, \tilde{\sigma}^{'}\}$ and
   \[
   \tilde{x}_{01}=[0 ~1], \tilde{x}_{m1}=[1~1],\tilde{x}_{02}=[1~0], \tilde{x}_{m2}=[1~1],
   \]
   \[\tilde{\sigma}_{1}=\left[
                                                \begin{array}{ccc}
                                                   1    & 0. 4 \\
                                                   0. 3  & 0. 5 \\
                                                \end{array}
                                             \right],
                                             \tilde{\sigma}_{1}^{'}=\left[
                                                \begin{array}{ccc}
                                                   0. 4  & 0. 7 \\
                                                   0. 7  & 1 \\
                                                \end{array}
                                             \right]
,
    \]
    \[
    \tilde{\sigma}_{2}=\left[
                                                \begin{array}{ccc}
                                                   0. 5  & 0. 3 \\
                                                   0. 3  & 1 \\
                                                \end{array}
                                             \right],
                                             \tilde{\sigma}_{2}^{'}=\left[
                                                \begin{array}{ccc}
                                                   1  & 0. 7 \\
                                                   0. 7  & 0. 4 \\
                                                \end{array}
                                             \right],
\]
where $\tilde{\sigma}_{i}$ and $\tilde{\sigma}_{i}^{'}$ are the corresponding matrices of events $\tilde{\sigma}$ and $\tilde{\sigma}^{'}$  in $\tilde{G}_{i}$, respectively. $\tilde{G}_{1}$ and $\tilde{G}_{2}$ are shown as Fig. 2. \par
We have the fuzzy relation
                            \[
                                \tilde{\phi}=\left[
                                                \begin{array}{ccc}
                                                   0. 5  & 1 \\
                                                   1  & 0. 5 \\
                                                \end{array}
                                             \right],
                            \]
 which satisfies the fuzzy simulation conditions with respect to $\tilde{G}_1\rightarrow\tilde{G}_2$ and $\tilde{G}_2\rightarrow\tilde{G}_1$. That is, $\tilde{G}_{1} \subseteq_{\tilde{\phi}} \tilde{G}_{2}$ and $\tilde{G}_{2} \subseteq_{\tilde{\phi}} \tilde{G}_{1}$. Therefore $\tilde{G}_{1}$ and $\tilde{G}_{2}$ are fuzzy simulation equivalent (as shown in Fig. 2).
\end{example}
\begin{figure}
\centering
\includegraphics[width=0.32\textwidth]{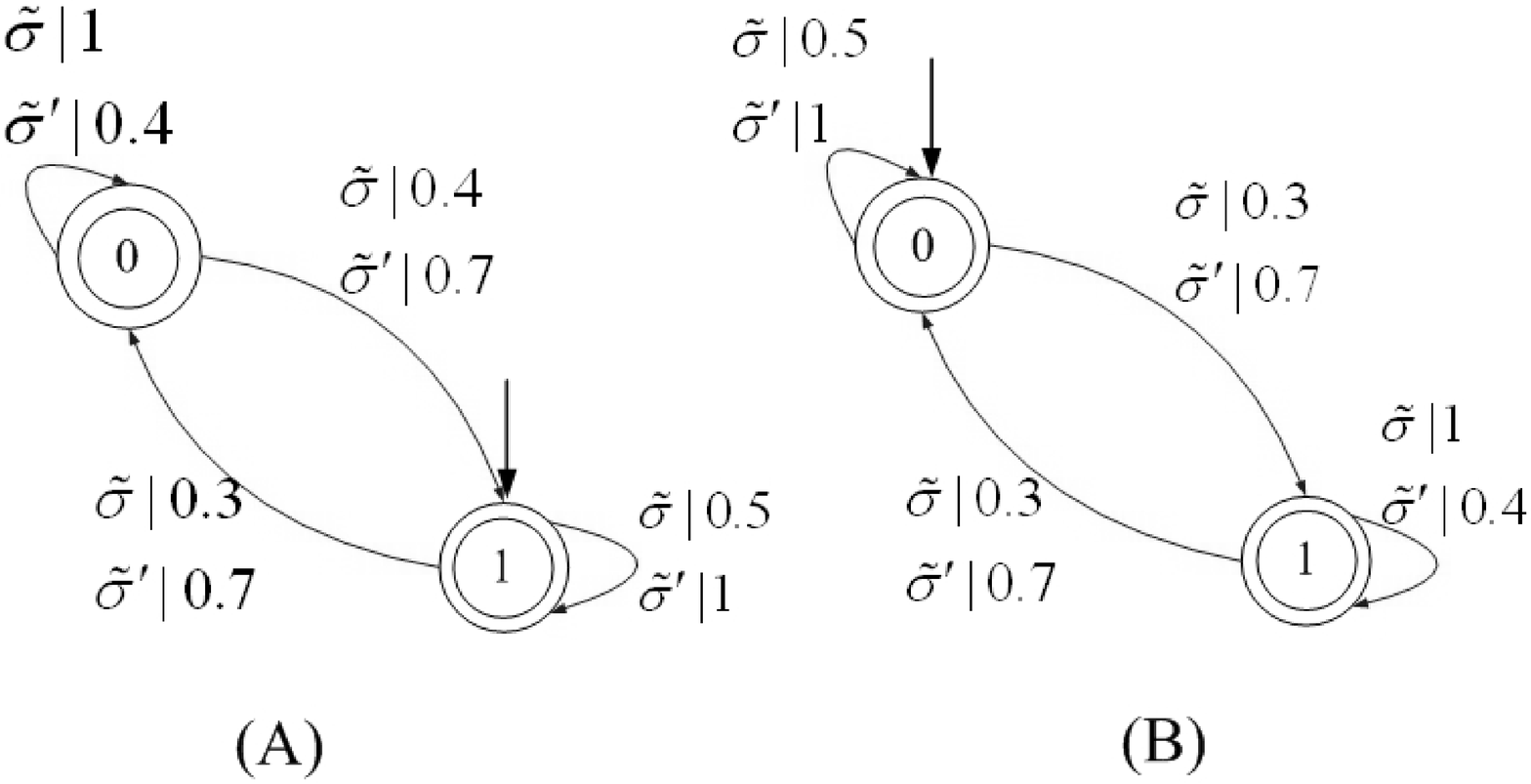}
\caption{\label{fig2} (A). FDES  $\tilde{G_{1}}$ of Example2.  (B). FDES $\tilde{G_{2}}$ of Example2. According to Definition 4 and Definition 5, $\tilde{G}_{1}\sim \tilde{G}_{2}$.}
\end{figure}

\subsection{Verification of Fuzzy Simulation Relations}
In this subsection, we present a method for deciding whether there is a fuzzy simulation between given FDESs.\par
From Definition 4, we notice that the verification of the fuzzy simulation relation means finding one of the solutions of  Equations (5a)-(5c).

Before giving the method for verifying the fuzzy simulation relation between two FDESs, we need to present a useful lemma as follows.\par
\begin{lemma}
Let two FDESs $\tilde{G}_{i}=\{ \tilde{X}_{i}, \tilde{\Sigma}, \tilde{\alpha}_{i}, \tilde{x}_{0i}, \tilde{x}_{mi}\}$, where $i\in\{1,2\}$, $|X_{1}| = m$, $|X_{2}|=n, |\tilde{\Sigma}| = k$. Suppose $A = \{\tilde{x}_{01,i},\tilde{x}_{02,j},\tilde{x}_{m1,i},\tilde{x}_{m2,j},\tilde{\sigma}_{1,ii^{'}},\tilde{\sigma}_{2,jj^{'}} \}$, where $i,i^{'} \in [1,m]$, $j,j^{'}\in[1,n]$, where $\tilde{\sigma}_{1},\tilde{\sigma}_{2}$ are the corresponding matrices for any event $\tilde{\sigma} \in \tilde{\Sigma}$ in $\tilde{G}_{1}$ and $\tilde{G}_{2}$, respectively. That is, $A$ is the set of all the elements in the state vectors and event matrices of $\tilde{G}_{i}$.
Let $\tilde{\phi}=[\tilde{\phi}_{ij}]_{m*n}$, $\tilde{\phi^{\uparrow}}=[\tilde{\phi^{\uparrow}}_{ij}]_{m*n}$, and  $\tilde{\phi^{\uparrow}}_{ij} = \min\{ a | a \in A \wedge a \geq \tilde{\phi}_{ij}  \} $.
If $\tilde{\phi}$ satisfies Equations (5a)-(5c), then $\tilde{\phi^{\uparrow}}$ also satisfies Equations (5a)-(5c).
\end{lemma}
\begin{IEEEproof}
\begin{enumerate}
\item
 It is obvious that $\tilde{\phi}\leq \tilde{\phi^{\uparrow}}$, which together with $\tilde{x}_{01} \leq  \tilde{x}_{02}$ $\odot \tilde{\phi}^{T}$ implies that $\tilde{x}_{01} \leq  \tilde{x}_{02}\odot \tilde{\phi^{\uparrow}}^{T}$.
\item
 $\tilde{x}_{m1}\odot \tilde{\phi} \leq \tilde{x}_{m2}$ means that $\forall j\in[1,n], \max_{i\in [1,m]}$ $\{\min\{\tilde{x}_{m1,i} , \tilde{\phi}_{ij}\}\} \leq \tilde{x}_{m2,j}$.
 That is, $\forall i\in[1,m],j\in[1,j]$, $\min\{\tilde{x}_{m1,i}, \tilde{\phi}_{ij}\}\leq \tilde{x}_{m2,j}$. Next we show that $\min\{\tilde{x}_{m1,i}, \tilde{\phi}^{\uparrow}_{ij}\}\leq \tilde{x}_{m2,j}$ holds by dividing into the following two cases:\par
 \begin{enumerate}
 \item
 ~when $\tilde{x}_{m1,i} \leq \tilde{\phi}_{ij} $,
$\min_{i\in [1,m]}\{\tilde{x}_{m1,i} , \tilde{\phi}^{\uparrow}_{ij}\}\leq \tilde{x}_{m2,j}$ is obvious.
\item
~when $\tilde{x}_{m1,i} \geq \tilde{\phi}_{ij} $,
we immediately get $\tilde{\phi}_{ij} \leq \tilde{x}_{m2,j} $.
For $\tilde{x}_{m2,j} \in A $, we get $\tilde{\phi}^{\uparrow}_{ij} \leq \tilde{x}_{m2,j}$.
Then $\min\{\tilde{x}_{m1,i}, \tilde{\phi}^{\uparrow}_{ij}\}\leq \tilde{x}_{m2,j}$ holds. \par
 \end{enumerate}
Therefore, $\max_{i\in [1,m]}\{\min\{\tilde{x}_{m1,i} , \tilde{\phi}^{\uparrow}_{ij}\}\}\leq \tilde{x}_{m2,j}$ holds for any $ j\in[1,n]$. That is, $\tilde{x}_{m1}\odot \tilde{\phi^{\uparrow}} \leq \tilde{x}_{m2}$.
\item
$\tilde{\phi}^{T} \odot \tilde{\sigma}_{1} \leq \tilde{\sigma}_{2}\odot \tilde{\phi}^{T}$ means that $\forall i \in [1,n],j \in [1,m]$,
\begin{multline*}
  \max_{j^{'} \in [1,n]}\{\min \{\tilde{\phi}^{T}_{ij^{'}},\tilde{\sigma}_{1,j^{'}j}\} \} \leq \\
   \max_{i^{'} \in [1,m]}\{\min \{ \tilde{\sigma}_{2,ii^{'}}, \tilde{\phi}^{T}_{i^{'}j}  \} \}.
\end{multline*}
Suppose when $i^{'} = i^{0}$, the right-hand side of the above inequality gets the maximum. Then  $\forall i \in [1,n],j \in [1,m],j^{'} \in [1,m]$, we get
 \[ \min \{\tilde{\phi}^{T}_{ij^{'}},\tilde{\sigma}_{1,j^{'}j}\}  \leq   \min \{ \tilde{\sigma}_{2,ii^{0}}, \tilde{\phi}^{T}_{i^{0}j}  \}. \]
Similarly, by dividing into $\tilde{\phi}^{T}_{ij^{'}} \leq \tilde{\sigma}_{1,j^{'}j}$ and $\tilde{\phi}^{T}_{ij^{'}} \geq \tilde{\sigma}_{1,j^{'}j}$ two cases, we have
\begin{equation}
\min \{\tilde{\phi^{\uparrow}}^{T}_{ij^{'}},\tilde{\sigma}_{1,j^{'}j}\}  \leq   \min \{ \tilde{\sigma}_{2,ii^{0}}, \tilde{\phi}^{T}_{i^{0}j}  \}.
\end{equation}
  Together with $\tilde{\phi}^{T}_{i^{0}j} \leq \tilde{\phi^{\uparrow}}^{T}_{i^{0}j} $, we further get
 \begin{equation}
 \min \{\tilde{\phi^{\uparrow}}^{T}_{ij^{'}},\tilde{\sigma}_{1,j^{'}j}\}  \leq   \min \{ \tilde{\sigma}_{2,ii^{0}}, \tilde{\phi^{\uparrow}}^{T}_{i^{0}j}  \}.
 \end{equation}
Then from Equations (6) and (7), we get that $\forall i \in [1,n],j \in [1,m]$,
\begin{multline*}
  \max_{j^{'} \in [1,n]}\{\min \{\tilde{\phi^{\uparrow}}^{T}_{ij^{'}},\tilde{\sigma}_{1,j^{'}j}\} \} \leq \\
   \max_{i^{'} \in [1,m]}\{\min \{ \tilde{\sigma}_{2,ii^{'}}, \tilde{\phi^{\uparrow}}^{T}_{i^{'}j}  \} \}.
\end{multline*}
That is, $\tilde{\phi^{\uparrow}}^{T} \odot \tilde{\sigma}_{1} \leq \tilde{\sigma}_{2}\odot \tilde{\phi^{\uparrow}}^{T}$.

\end{enumerate}
Therefore, $\tilde{\phi^{\uparrow}}$ satisfies Equations (5a)-(5c).
\end{IEEEproof}
\par
\begin{remark}
Lemma 1 suggests that if there exists a fuzzy simulation relation $\tilde{\phi}$ between $\tilde{G}_{1}$ and $\tilde{G}_{2}$, then there also exists a fuzzy simulation relation $\tilde{\phi^{\uparrow}}$, whose elements are all from the set $A$, between $\tilde{G}_{1}$ and $\tilde{G}_{2}$. Equivalently, if there does not exist any fuzzy simulation relation whose elements are all from the set $A$, then there does not exist any  fuzzy simulation relation. In addition, the cardinality of set $A$ is finite. Thus, we can make an exhaustive search for the fuzzy simulation relations over the matrix space $A^{m*n}$. The worst case complexity of the search algorithm is $O((2*(m+n)+(m^2+n^2)*k)^{m*n})$.
Notably, Reference \cite{mc-2} provides an algorithm  to compute the greatest simulation between fuzzy automata. The algorithm can be used  for verifying simulation.
\end{remark}
\par
The following example illustrates the search algorithm to decide whether
there is a simulation between the given FDESs.

\begin{example}
   Let $\tilde{G}_{i}=\{\tilde{X}_{i}, \tilde{\Sigma}, \tilde{\alpha}_{i}, \tilde{x}_{0i}, \tilde{x}_{mi}\}, i\in\{1, 2\}$, where $\tilde{\Sigma}=\{\tilde{\sigma}, \tilde{\sigma}^{'}\}$ and
   \[
   \tilde{x}_{01}=[0. 9~1], \tilde{x}_{m1}=[1~1],\tilde{x}_{02}=[0~0~1], \tilde{x}_{m2}=[1~1~1],
   \]
   \[
   \tilde{\sigma}_{1}=\left[
                                                \begin{array}{ccc}
                                                   1    & 0. 7 \\
                                                   0. 7  & 0. 9 \\
                                                \end{array}
                                             \right],
                                             \tilde{\sigma}_{1}^{'}=\left[
                                                \begin{array}{ccc}
                                                   0. 7  & 0. 7 \\
                                                   0. 9  & 1 \\
                                                \end{array}
                                             \right],
 \]
   \[
    \tilde{\sigma}_{2}=\left[
                                                \begin{array}{ccc}
                                                   1  & 0. 2 & 0. 4\\
                                                   0. 6  & 1 & 0. 2\\
                                                   0. 4 &  0. 7 & 0. 9 \\
                                                \end{array}
                                             \right],
                                             \tilde{\sigma}_{2}^{'}=\left[
                                                \begin{array}{ccc}
                                                   0. 6  & 0. 7 & 0. 1 \\
                                                   0. 7  & 0. 2 & 0. 4\\
                                                   0. 9 & 0. 9 & 1 \\
                                                \end{array}
                                             \right].
\]
Here $\tilde{\sigma}_{i}$ and $\tilde{\sigma}_{i}^{'}$ are the corresponding matrices of events $\tilde{\sigma}$ and $\tilde{\sigma}^{'}$ in $\tilde{G}_{i}$, respectively. \par
First, we compute the set
 $ A = \{0 , 0.2 , 0.4 , 0.7 , 0.9 , 1 \}$. By exhaustive searching over the matrix space $A^{2*3}$ and $A^{3*2}$, respectively, we get the following fuzzy simulation relations
                            \[
                                \tilde{\phi}=\left[
                                                \begin{array}{ccc}
                                                   1  & 1 & 0. 9 \\
                                                   0. 7  & 0. 7 & 1\\
                                                \end{array}
                                             \right] ~\text{and}~
                              \tilde{\varphi}=\left[
                                                \begin{array}{ccc}
                                                   1  & 0. 9  \\
                                                   1  & 0. 9 \\
                                                   0. 7 & 1  \\
                                                \end{array}
                                             \right],
                            \]
such that $\tilde{G}_{1} \subseteq_{\tilde{\phi}} \tilde{G}_{2}$ and $\tilde{G}_{2} \subseteq_{\tilde{\varphi}} \tilde{G}_{1}$. Therefore, $\tilde{G}_{1}$ and $\tilde{G}_{2}$ are fuzzy simulation equivalent.
\end{example}
\subsection{Properties of Fuzzy Simulation of FDESs}
 In this subsection, we discuss some basic properties of fuzzy simulation, which play an important role in the study of fuzzy simulation-equivalence control of FDESs. For convenience, we consider a set of FDESs with common events $\mathcal{A}=\{\tilde{G}_{i}\}$, where $\tilde{G}_{i}=\{\tilde{X}_{i}, \tilde{\Sigma}, \tilde{\alpha}_{i}, \tilde{x}_{0i}, \tilde{x}_{mi}\}, i\in\{1, 2, 3, \ldots\}$.
\begin{proposition}
  The fuzzy simulation relation is transitive.
\end{proposition}
\begin{IEEEproof}
   Consider $\tilde{G}_{i}\in\mathcal{A}, i=\{1, 2, 3\}$ with $\tilde{G}_{1}\subseteq_{\tilde{\phi}_{1}} \tilde{G}_{2}$ and  $\tilde{G}_{2}\subseteq_{\tilde{\phi}_{2}} \tilde{G}_{3}$. We need to show that there exists a simulation relation $\tilde{\phi}$ such that $\tilde{G}_{1}\subseteq_{\tilde{\phi}} \tilde{G}_{3}$. \par
  \begin{enumerate}
    \item By $\tilde{G}_{2}\subseteq_{\tilde{\phi}_{2}} \tilde{G}_{3}$, we have $\tilde{x}_{02}\leq \tilde{x}_{03}\odot \tilde{\phi}_{2}^{T}$,
which implies $\tilde{x}_{02} \odot \tilde{\phi}_{1}^{T}  \leq \tilde{x}_{03}\odot \tilde{\phi}_{2}^{T} \odot \tilde{\phi}_{1}^{T}$. Noting that $\tilde{\phi}_{2}^{T} \odot \tilde{\phi}_{1}^{T} = (\tilde{\phi}_{1} \odot \tilde{\phi}_{2})^{T}$, we get $\tilde{x}_{02} \odot \tilde{\phi}_{1}^{T}  \leq \tilde{x}_{03}\odot (\tilde{\phi}_{1} \odot \tilde{\phi}_{2})^{T} $. By $\tilde{G}_{1}\subseteq_{\tilde{\phi}_{1}} \tilde{G}_{2}$, we have $\tilde{x}_{01}\leq \tilde{x}_{02}\odot \tilde{\phi}_{1}^{T}$. Then we get $\tilde{x}_{01}\leq \tilde{x}_{03}\odot (\tilde{\phi}_{1} \odot \tilde{\phi}_{2})^{T}$. \par
    \item By  $\tilde{G}_{1}\subseteq_{\tilde{\phi}_{1}} \tilde{G}_{2}$, we have $\tilde{x}_{m1}\odot \tilde{\phi}_{1} \leq \tilde{x}_{m2}$, which implies $\tilde{x}_{m1}\odot \tilde{\phi}_{1}\odot \tilde{\phi}_{2} \leq \tilde{x}_{m2} \odot \tilde{\phi}_{2}$. By $\tilde{G}_{2}\subseteq_{\tilde{\phi}_{2}} \tilde{G}_{3}$, we get $\tilde{x}_{m2}\odot \tilde{\phi}_{2} \leq \tilde{x}_{m3}$. Then we have $\tilde{x}_{m1}\odot (\tilde{\phi}_{1}\odot\tilde{\phi}_{2}) \leq \tilde{x}_{m3}$.\par
    \item By $\tilde{G}_{1}\subseteq_{\tilde{\phi}_{1}} \tilde{G}_{2}$ and  $\tilde{G}_{2}\subseteq_{\tilde{\phi}_{2}} \tilde{G}_{3}$,
 we have $\tilde{\phi}_{1}^{T} \odot \tilde{\sigma}_{1} \leq \tilde{\sigma}_{2}\odot \tilde{\phi}_{1}^{T}$ and $\tilde{\phi}_{2}^{T} \odot \tilde{\sigma}_{2} \leq \tilde{\sigma}_{3}\odot \tilde{\phi}_{2}^{T}$, which imply $\tilde{\phi}_{2}^{T} \odot \tilde{\phi}_{1}^{T} \odot \tilde{\sigma}_{1} \leq \tilde{\phi}_{2}^{T} \odot \tilde{\sigma}_{2}\odot \tilde{\phi}_{1}^{T}$ and  $\tilde{\phi}_{2}^{T} \odot \tilde{\sigma}_{2} \odot \tilde{\phi}_{1}^{T} \leq \tilde{\sigma}_{3}\odot \tilde{\phi}_{2}^{T}\odot \tilde{\phi}_{1}^{T}$, respectively.
 Thus we have $\tilde{\phi}_{2}^{T} \odot \tilde{\phi}_{1}^{T} \odot \tilde{\sigma}_{1} \leq \tilde{\sigma}_{3}\odot \tilde{\phi}_{2}^{T}\odot \tilde{\phi}_{1}^{T}$, that is, $(\tilde{\phi}_{1}\odot\tilde{\phi}_{2})^{T} \odot \tilde{\sigma}_{1} \leq \tilde{\sigma}_{3}\odot (\tilde{\phi}_{1}\odot \tilde{\phi}_{2})^{T}$. \par
  \end{enumerate}\par
 Hence, let $\tilde{\phi} = \tilde{\phi}_{1}\odot\tilde{\phi}_{2}$. By the definition of fuzzy simulation, we have $G_{1}\subseteq_{\tilde{\phi}} G_{3}$.
\end{IEEEproof}

\begin{lemma}
 Assume A, B, C, and D are matrices for which $A\odot C$ and $B\odot D$ are defined. Then
 $  (A \tilde{\otimes} B)\odot(C \tilde{\otimes} D) = (A\odot C) \tilde{\otimes} (B\odot D).$
\end{lemma}
\begin{IEEEproof}
  Without loss of generality, suppose $A, B, C, D$ are $k*m, p*s, m*n, s*r$ matrices respectively. Let $a\wedge B$ denote $Min\{a, B\}$ and let $a\wedge c$ denote $\min\{a, c\}$. Then we have\par
 \begin{align}
  &(A \tilde{\otimes} B)\odot(C \tilde{\otimes} D)\nonumber\\
 &=
  \left[ \begin{array}{ccc}
            a_{11}\wedge B & \ldots & a_{1m}\wedge B \\
            \vdots & \ddots & \vdots\\
            a_{k1}\wedge B & \ldots & a_{km}\wedge B
                        \end{array} \right]  \odot \nonumber \\
           & \left[ \begin{array}{ccc}
            c_{11}\wedge D & \ldots & c_{1n}\wedge D \\
            \vdots & \ddots & \vdots\\
            c_{m1}\wedge D & \ldots & c_{mn}\wedge D
            \end{array} \right] \nonumber\\
&= \left[\max_{l=1}^{m}\min\{(a_{il}\wedge B)\odot (c_{lj}\wedge D)\}\right]_{i\in[1, k]}^{j\in[1, n]} \nonumber\\
&= \left[\max_{l=1}^{m}\min\{(a_{il}\wedge c_{lj})\wedge (B\odot D)\}\right]_{i\in[1, k]}^{j\in[1, n]} \nonumber\\
&= \left[ [A \odot C ]_{ij} \wedge (B\odot D) \right]_{i\in[1, k]}^{j\in[1, n]}\nonumber\\
&= (A\odot C)\tilde{\otimes} (B\odot D). \nonumber
\end{align}
\end{IEEEproof}
\par
   Lemma 2 is used to support the proof of Proposition 2 and Proposition 3.
\par
The following proposition shows that if the first fuzzy automaton can be simulated by the second automaton, then the parallel composition
of the first automaton and another automaton can also be simulated by the second automaton.

 \begin{proposition}
  $\tilde{G}_{1}\subseteq \tilde{G}_{3}\Rightarrow \tilde{G}_{1}|| \tilde{G}_{2} \subseteq \tilde{G}_{3}$  ; $\tilde{G}_{2}\subseteq \tilde{G}_{3}\Rightarrow \tilde{G}_{1}|| \tilde{G}_{2} \subseteq \tilde{G}_{3}$.
\end{proposition}
\begin{IEEEproof}
We prove the part 1 first. Suppose $\tilde{G}_{1}\subseteq_{\tilde{\phi}_{1}} \tilde{G}_{3}$ and $|X_{2}| = n$. We define $\tilde{\phi}_{2}:= \tilde{\phi}_{1} \tilde{\otimes} (\textbf{1})_{n}^{T}$, in which $(\textbf{1})_{n}=[\underbrace{1, \ldots, 1}_{n}]$. We show
$\tilde{G}_{1}|| \tilde{G}_{2} \subseteq_{\tilde{\phi}_{2}} \tilde{G}_{3}$ as follows.\par
\begin{enumerate}
  \item  By $\tilde{G}_{1}\subseteq_{\tilde{\phi}_{1}} \tilde{G}_{3}$, we have $\tilde{x}_{01}\leq \tilde{x}_{03} \odot \tilde{\phi}_{1}^{T}$, which implies $\tilde{x}_{01} \tilde{\otimes} (\textbf{1})_{n} \leq \tilde{x}_{03} \odot \tilde{\phi}_{1}^{T} \tilde{\otimes} (\textbf{1})_{n}$. As $\tilde{x}_{01} \tilde{\otimes} \tilde{x}_{02} \leq \tilde{x}_{01} \tilde{\otimes} (\textbf{1})_{n}$ is obvious,  we have $\tilde{x}_{01} \tilde{\otimes} \tilde{x}_{02} \leq \tilde{x}_{03} \odot \tilde{\phi}_{1}^{T} \tilde{\otimes} (\textbf{1})_{n} $. Noting that a matrix's $\tilde{\otimes}$ operation with $ (\textbf{1})_{n}$ just means successively duplicating its every column $n$ times, we get $\tilde{x}_{03} \odot \tilde{\phi}_{1}^{T} \tilde{\otimes} (\textbf{1})_{n} = \tilde{x}_{03} \odot (\tilde{\phi}_{1}^{T} \tilde{\otimes} (\textbf{1})_{n})$. Therefore, we have $\tilde{x}_{01} \tilde{\otimes} \tilde{x}_{02} \leq \tilde{x}_{03} \odot (\tilde{\phi}_{1}^{T} \tilde{\otimes} (\textbf{1})_{n}) = \tilde{x}_{03} \odot \tilde{\phi}_{2}^{T}$. \par
  \item As $(\tilde{x}_{m2} \odot (\textbf{1})_{n}^{T})$ is a 1*1's matrix, we have $(\tilde{x}_{m1}\odot\tilde{\phi}_{1})\tilde{\otimes}(\tilde{x}_{m2}\odot (\textbf{1})_{n}^{T}) \leq (\tilde{x}_{m1}\odot\tilde{\phi}_{1}) $.
      By Lemma 2, we have $(\tilde{x}_{m1}\odot\tilde{\phi}_{1})\tilde{\otimes}(\tilde{x}_{m2}\odot (\textbf{1})_{n}^{T}) = (\tilde{x}_{m1} \tilde{\otimes} \tilde{x}_{m2}) \odot (\tilde{\phi}_{1}\tilde{\otimes} (\textbf{1})_{n}^{T} )$.
      Then we have $(\tilde{x}_{m1} \tilde{\otimes} \tilde{x}_{m2}) \odot (\tilde{\phi}_{1}\tilde{\otimes} (\textbf{1})_{n}^{T} ) \leq (\tilde{x}_{m1}\odot\tilde{\phi}_{1}) $. Further, by $\tilde{G}_{1}\subseteq_{\tilde{\phi}_{1}} \tilde{G}_{3}$, we have $\tilde{x}_{m1}\odot \tilde{\phi}_{1}\leq \tilde{x}_{m3}$. Therefore $(\tilde{x}_{m1} \tilde{\otimes} \tilde{x}_{m2}) \odot \tilde{\phi}_{2} = (\tilde{x}_{m1} \tilde{\otimes} \tilde{x}_{m2}) \odot (\tilde{\phi}_{1}\tilde{\otimes} (\textbf{1})_{n}^{T} )\leq \tilde{x}_{m3}$ holds. \par

  \item By $\tilde{G}_{1}\subseteq_{\tilde{\phi}_{1}} \tilde{G}_{3}$, we get $\tilde{\phi}_{1}^{T} \odot \tilde{\sigma}_{1} \leq \tilde{\sigma}_{3}\odot \tilde{\phi}_{1}^{T}$, which together with $(\textbf{1})_{n}\odot \tilde{\sigma}_{2} \leq (\textbf{1})_{n}$ implies $ (\tilde{\phi}_{1}^{T} \odot \tilde{\sigma}_{1}) \tilde{\otimes} ((\textbf{1})_{n}\odot \tilde{\sigma}_{2}) \leq (\tilde{\sigma}_{3}\odot \tilde{\phi}_{1}^{T}) \tilde{\otimes} (\textbf{1})_{n}$.
Further, $(\tilde{\phi}_{1}^{T} \odot \tilde{\sigma}_{1}) \tilde{\otimes} ((\textbf{1})_{n}\odot \tilde{\sigma}_{2}) = (\phi_{1}^{T} \tilde{\otimes} (\textbf{1})_{n}) \odot (\tilde{\sigma}_{1}\tilde{\otimes} \tilde{\sigma}_{2}) $ holds by Lemma 2 and $(\tilde{\sigma}_{3}\odot \tilde{\phi}_{1}^{T}) \tilde{\otimes} (\textbf{1})_{n} =\tilde{\sigma}_{3}\odot (\tilde{\phi}_{1}^{T} \tilde{\otimes} (\textbf{1})_{n} )$ holds as we interpreted above. Therefore, we get $(\tilde{\phi}_{1}^{T} \tilde{\otimes} (\textbf{1})_{n}) \odot (\tilde{\sigma}_{1}\tilde{\otimes} \tilde{\sigma}_{2}) \leq \tilde{\sigma}_{3}\odot (\tilde{\phi}_{1}^{T} \tilde{\otimes} (\textbf{1})_{n}) $, that is, $\tilde{\phi}_{2} \odot (\tilde{\sigma}_{1}\tilde{\otimes} \tilde{\sigma}_{2}) \leq \tilde{\sigma}_{3}\odot \tilde{\phi}_{2} $.
\end{enumerate}\par
That is, we complete the proof of part 1 of the proposition. \par
Similarly, supposing $\tilde{G}_{2}\subseteq_{\tilde{\phi}_{1}} \tilde{G}_{3}$, $|X_{1}| = n$, and defining $\tilde{\phi}_{2} := (\textbf{1})_{n}^{T} \tilde{\otimes} \tilde{\phi}_{1} $, we can prove $\tilde{G}_{2}\subseteq_{\tilde{\phi}_{1}} \tilde{G}_{3}\Rightarrow \tilde{G}_{1}|| \tilde{G}_{2} \subseteq_{\tilde{\phi}_{2}} \tilde{G}_{3}$.
\end{IEEEproof}
\par
The following corollary follows from Proposition 2.
\begin{corollary}
  $\tilde{G}_{1}||\tilde{G}_{2}\subseteq \tilde{G}_{1} ; \tilde{G}_{1}||\tilde{G}_{2}\subseteq \tilde{G}_{2}$.
\end{corollary}
\begin{IEEEproof}
   Since $\tilde{G}_{1}\subseteq \tilde{G}_{1}$ and $\tilde{G}_{2}\subseteq \tilde{G}_{2}$, by Proposition 2, we immediately get $\tilde{G}_{1}||\tilde{G}_{2}\subseteq \tilde{G}_{1}$ and $\tilde{G}_{1}||\tilde{G}_{2}\subseteq \tilde{G}_{2}$.
\end{IEEEproof}

\begin{proposition}
  Given two fuzzy automata $\tilde{G}_{i}, i\in\{1, 2\}$, then $L(\tilde{G}_{1}||\tilde{G}_{2}) = L(\tilde{G}_{1})\tilde{\cap} L(\tilde{G}_{2})$, where symbol $\tilde{\cap}$ is Zadeh fuzzy AND operator.
\end{proposition}
\begin{IEEEproof}
 Let $|X_{1}|=m$ and $|X_{2}|=n$. Suppose for any $\tilde{s}\in\tilde{\Sigma}^{*}$ with $\tilde{s} = \tilde{\sigma}^{1}\tilde{\sigma}^{2}\ldots\tilde{\sigma}^{k}$, the corresponding matrices of fuzzy event $\tilde{\sigma}^{i}, i\in[1, k]$ in $\tilde{G}_{1}$ and $\tilde{G}_{2}$  are denoted by $\tilde{\sigma}^{i}_{1}$ and $\tilde{\sigma}^{i}_{2}$, respectively. For convenience, let $ (\tilde{\sigma}^{1}_{1}\odot\tilde{\sigma}^{2}_{1}\odot \ldots\odot\tilde{\sigma}^{k}_{1}) =  \tilde{\sigma}^{s}_{1}$ and $(\tilde{\sigma}^{1}_{1}\odot\tilde{\sigma}^{2}_{1}\odot\ldots\odot\tilde{\sigma}^{k}_{1}) = \tilde{\sigma}^{s}_{2} $. We have:
\begin{align}
&L(\tilde{G}_{1}||\tilde{G}_{2})(\tilde{s})\nonumber\\
&= \max_{i=1}^{m*n}(\tilde{x}_{01}\tilde{\otimes} \tilde{x}_{02})\odot (\tilde{\sigma}^{s}_{1} \tilde{\otimes} \tilde{\sigma}^{s}_{2})*\bar{s}_{i}\nonumber\\
&=\max_{i=1}^{m*n}(\tilde{x}_{01} \odot \tilde{\sigma}^{s}_{1})  \tilde{\otimes}( \tilde{x}_{02} \odot  \tilde{\sigma}^{s}_{2})*\bar{s}_{i}\nonumber\\
&=\min\{ \max_{i=1}^{m}({\tilde{x}}_{01}\odot{\tilde{\sigma}}^{s}_{1}*\bar{s}_{i}), \max_{i=1}^{n}(\tilde{x}_{02} \odot {\tilde{\sigma}}^{s}_{1}*\bar{s}_{i}) \}\nonumber\\
&=\min\{L(\tilde{G}_{1})(\tilde{s}), L(\tilde{G}_{2})(\tilde{s})\}\nonumber\\
&=L(\tilde{G}_{1})(\tilde{s}) \tilde{\cap} L(\tilde{G}_{2})(\tilde{s}).\nonumber
\end{align}
\end{IEEEproof}
\par
The following proposition shows that if the first fuzzy automaton can be fuzzy simulated by another two automata, then the first automaton also can be fuzzy simulated by the parallel composition of another two automata.
 \begin{proposition}
  $\tilde{G}_{3}\subseteq \tilde{G}_{1}, \tilde{G}_{3} \subseteq \tilde{G}_{2} \Rightarrow \tilde{G}_{3} \subseteq \tilde{G}_{1}||\tilde{G}_{2}$.
\end{proposition}
\begin{IEEEproof}
 We would like to postpone the proof to Appendix A.
\end{IEEEproof}
\par
 The following proposition shows that the inverse direction of Proposition 4 also holds.
\begin{proposition}
  $\tilde{G}_{3} \subseteq \tilde{G}_{1}||\tilde{G}_{2} \Rightarrow\tilde{G}_{3}\subseteq \tilde{G}_{1}, \tilde{G}_{3} \subseteq \tilde{G}_{2}$
\end{proposition}
\begin{IEEEproof}
 We also would like to postpone the proof to  Appendix B.
\end{IEEEproof}
\par
The following corollary follows from Proposition 2 and Proposition 4.
\begin{corollary}
  $\tilde{G}_{1}\subseteq \tilde{G}_{2} \Rightarrow \tilde{G}_{3}||\tilde{G}_{1} \subseteq \tilde{G}_{3}||\tilde{G}_{2}$.
\end{corollary}
\begin{IEEEproof}
 Since $\tilde{G}_{1}\subseteq \tilde{G}_{2}$, by Proposition 2, we have $\tilde{G}_{3}||\tilde{G}_{1} \subseteq \tilde{G}_{2}$. By Corollary 1, we  have $\tilde{G}_{3}||\tilde{G}_{1} \subseteq \tilde{G}_{3}$. Therefore by Proposition 4, we further have $\tilde{G}_{3}||\tilde{G}_{1} \subseteq \tilde{G}_{3}||\tilde{G}_{2}$.
\end{IEEEproof}

\section{Fuzzy Simulation-Equivalence control of FDESs}

In this section, we first study the fuzzy simulation-equivalence control problem, then investigate the relations between fuzzy language-equivalence control and fuzzy simulation-equivalence control.  \par

\subsection{Fuzzy Simulation-Equivalence Control}
We model an uncontrolled system, a specification, and a supervisor as the following fuzzy automata: $\tilde{G}=\{\tilde{X},\tilde{\Sigma},$ $ \tilde{\alpha}, \tilde{x}_{0}, \tilde{x}_{m}\}$, $\tilde{R}=\{\tilde{Q}, \tilde{\Sigma}, \tilde{\beta}, \tilde{q}_{0}, \tilde{q}_{m}\}$, and $\tilde{S}=\{\tilde{Y}, \tilde{\Sigma}, \tilde{\gamma}, $ $\tilde{y}_{0}, \tilde{y}_{m}\}$, respectively. In this subsection, we study the fuzzy simulation-equivalence control problem of FDESs, which guarantees the fuzzy simulation equivalence of the controlled system and the given specification, that is, $\tilde{G} || \tilde{S} \sim \tilde{R}$.\par

In an FDES, each fuzzy event is physically associated with a degree of uncontrollability. More formally, we present the following definition.
\begin{definition}
  The uncontrollable event set $\tilde{\Sigma}_{uc}$ and controllable event set $\tilde{\Sigma}_{c}$ are, respectively, defined as a function from $\tilde{\Sigma}$ to $[0, 1]$, which satisfy the following condition:
  \begin{equation}
  \tilde{\Sigma}_{uc}(\tilde{\sigma})+\tilde{\Sigma}_{c}(\tilde{\sigma}) = 1 ~~(\forall \tilde{\sigma} \in \tilde{\Sigma}),
  \end{equation}
  where  $\tilde{\Sigma}_{uc}(\tilde{\sigma})$ and $\tilde{\Sigma}_{c}(\tilde{\sigma})$ are the degrees of uncontrollability and controllability, respectively, of event $\tilde{\sigma}$.
\end{definition}
\par
Due to the uncontrollability of fuzzy event, we present the following notion to characterize the valid supervisors of fuzzy simulation-equivalence control.
\begin{definition}
    A fuzzy automaton $\tilde{S}= \{\tilde{Y}, \tilde{\Sigma}, \tilde{\gamma}, \tilde{y}_{0}, \tilde{y}_{m}\}$ with uncontrollable event set $\tilde{\Sigma}_{uc}$ and $|Y| = n$, is called a \emph{fuzzy $\tilde{\Sigma}_{u}$-compatible supervisor} if the following condition holds:
    \begin{equation}
        \max_{j=1}^{n}\tilde{\sigma}(i)(j) \geq \tilde{\Sigma}_{uc}(\tilde{\sigma}) ~(\forall \tilde{\sigma}\in\tilde{\Sigma}, ~\forall ~ i\in[1, n]).
    \end{equation}
\end{definition}
    \par
Intuitively, Equation (9) indicates that every row of every event matrix of the fuzzy $\tilde{\Sigma}_{u}$-compatible supervisor includes at least one element which is no less than the uncontrollable degree of the corresponding event.
\begin{remark}
\emph{Fuzzy $\tilde{\Sigma}_{u}$-compatible supervisor} generalizes the notion of \emph{$\tilde{\Sigma}_{u}$-compatible supervisor} introduced in \cite{zhou-1}. If we assume that the events, the states and the uncontrollability are all crisp, then it reduces to the $\tilde{\Sigma}_{u}$-compatible supervisor.
\end{remark}
\par
Next, we consider to find a necessary and sufficient condition for the existence of fuzzy supervisors. Intuitively, we believe that the fuzzy supervisor should be closely related to the specification ${\tilde{R}}$. Therefore, firstly we
construct a fuzzy $\tilde{\Sigma}_{uc}$-compatible supervisor ${\tilde{R}^{+}}$ based on the specification automaton $R$ as follows.
\begin{algorithm}
Supposing $\tilde{R}=\{\tilde{Q}, \tilde{\Sigma}, \tilde{\beta}, \tilde{q}_{0}, \tilde{q}_{m}\}$, $|Q| = n $, then we define
  \[
   \tilde{R}^{+}=\{\tilde{Q^{+}}, \tilde{\Sigma}^{+}, \tilde{\beta}^{+}, \tilde{q}_{0}^{+}, \tilde{q}_{m}^{+}  \},
   \]
   where $Q^{+} = Q \cup \{q^{+}\}$, $\tilde{q}_{0}^{+} = [\tilde{q}_{0}, 0]$, $\tilde{q}_{m}^{+}=[\tilde{q}_{m}, 0]$, $\tilde{\beta}^{+}:\tilde{Q^{+}} \times \tilde{\Sigma}^{+}\rightarrow \tilde{Q^{+}}$  is a transition function which is defined by $\tilde{\beta}^{+}(\tilde{q}, \tilde{\sigma} )=\tilde{q} \odot \tilde{\sigma}$ for $\tilde{q}\in \tilde{Q^{+}}$ and $\tilde{\sigma} \in \tilde{\Sigma}^{+} $. As the number of the corresponding crisp states increases to $(n+1)$, the order of the events matrices should increase to $(n+1)$. For any $ \tilde{\sigma}^{+}\in\tilde{\Sigma}^{+}$ and $\forall i, j\in[1, n+1]$, we construct it as follows:
   \begin{equation}
\tilde{\sigma}^{+}(i)(j) =
    \begin{cases}
    \tilde{\sigma}(i)(j),                                     & \text{if}\ \ \ \ \ \  i\in[1, n], j\in[1, n],\\
    0,                                                    & \text{if} \ \ \ \ \ \ i=n+1,j\in[1, n],\\
    \tilde{\Sigma}_{uc}(\tilde{\sigma}),                    & \text{if}\ \ \ \ \ \  i=n+1,j=n+1,\\
    0,                         & \text{else if}~ \max_{j=1}^{n}\tilde{\sigma}(i)(j) \geq \tilde{\Sigma}_{uc}(\tilde{\sigma}),\\
    \tilde{\Sigma}_{uc}(\tilde{\sigma}),  & \text{else if}~ \max_{j=1}^{n}\tilde{\sigma}(i)(j) < \tilde{\Sigma}_{uc}(\tilde{\sigma}).\\
    \end{cases}
\end{equation}\par
Here $\tilde{\sigma}(i)(j)$ denotes the $i$th row and $j$th column element of the matrix $\tilde{\sigma}$.
\end{algorithm}
\par
  The algorithm shows that $\tilde{R}^{+}$ is obtained by adding a crisp state and adding transitions from each state to the new state to ensure that $\tilde{R}^{+}$ is a fuzzy  $\tilde{\Sigma}_{uc}$-compatible supervisor. The following example illustrates the algorithm.
\begin{example}
      Let the specification $\tilde{R}=\{\tilde{Q}, \tilde{\Sigma}, \tilde{\beta}, \tilde{q}_{0}, \tilde{q}_{m}\}$.
   Here $\tilde{q}_{0}=[1~0]$,  $\tilde{q}_{m}=[0~1]$, $\tilde{\Sigma}=\{\tilde{\sigma}, \tilde{\sigma}^{'}\}$ , $\tilde{\Sigma}_{uc}(\tilde{\sigma})=0. 7$ and $\tilde{\Sigma}_{uc}(\tilde{\sigma}^{'})=0. 6$. The corresponding events matrices are:
    \[
                                 \tilde{\sigma}=\left[
                                                \begin{array}{cc}
                                                   0.8 & 0.4 \\
                                                   0.3   & 0 \\
                                                \end{array}
                                             \right],
                                             \tilde{\sigma}^{'}=\left[
                                                \begin{array}{ccc}
                                                   0 & 0.5 \\
                                                   0.3 & 0.7\\
                                               \end{array}
                                             \right].
   \]
Then by  Algorithm 1,  $\tilde{R}^{+}=\{\tilde{Q^{+}}, \tilde{\Sigma}^{+}, \tilde{\beta}^{+}, \tilde{q}_{0}^{+}, \tilde{q}_{m}^{+}  \}$.
Here $\tilde{q}_{0}^{+} = [1,0,0]$, $\tilde{q}_{m}^{+}=[0,1,0]$, and the corresponding events matrices are:
    \[
                                 \tilde{\sigma}=\left[
                                                \begin{array}{ccc}
                                                   0.8 & 0.4 & 0 \\
                                                   0.3   & 0 & 0.7\\
                                                   0     &  0 & 0.7\\
                                                \end{array}
                                             \right],
                                             \tilde{\sigma}^{'}=\left[
                                                \begin{array}{ccc}
                                                   0 & 0.5 & 0.6\\
                                                   0.3 & 0.7 & 0\\
                                                   0 & 0 & 0.6\\
                                               \end{array}
                                             \right].
   \]
The FDES $\tilde{R}$ and $\tilde{R}^{+}$ are shown as Fig. 3 (A) and (B), respectively.
\end{example}

\par
The following two lemmas characterize the relations among $\tilde{R}^{+}$, $\tilde{R}$, and any fuzzy $\tilde{\Sigma}_{uc}$-compatible supervisor $\tilde{S}$. They will be used to support the proof of Theorem 1.

\begin{figure}
\centering
\includegraphics[width=0.32\textwidth]{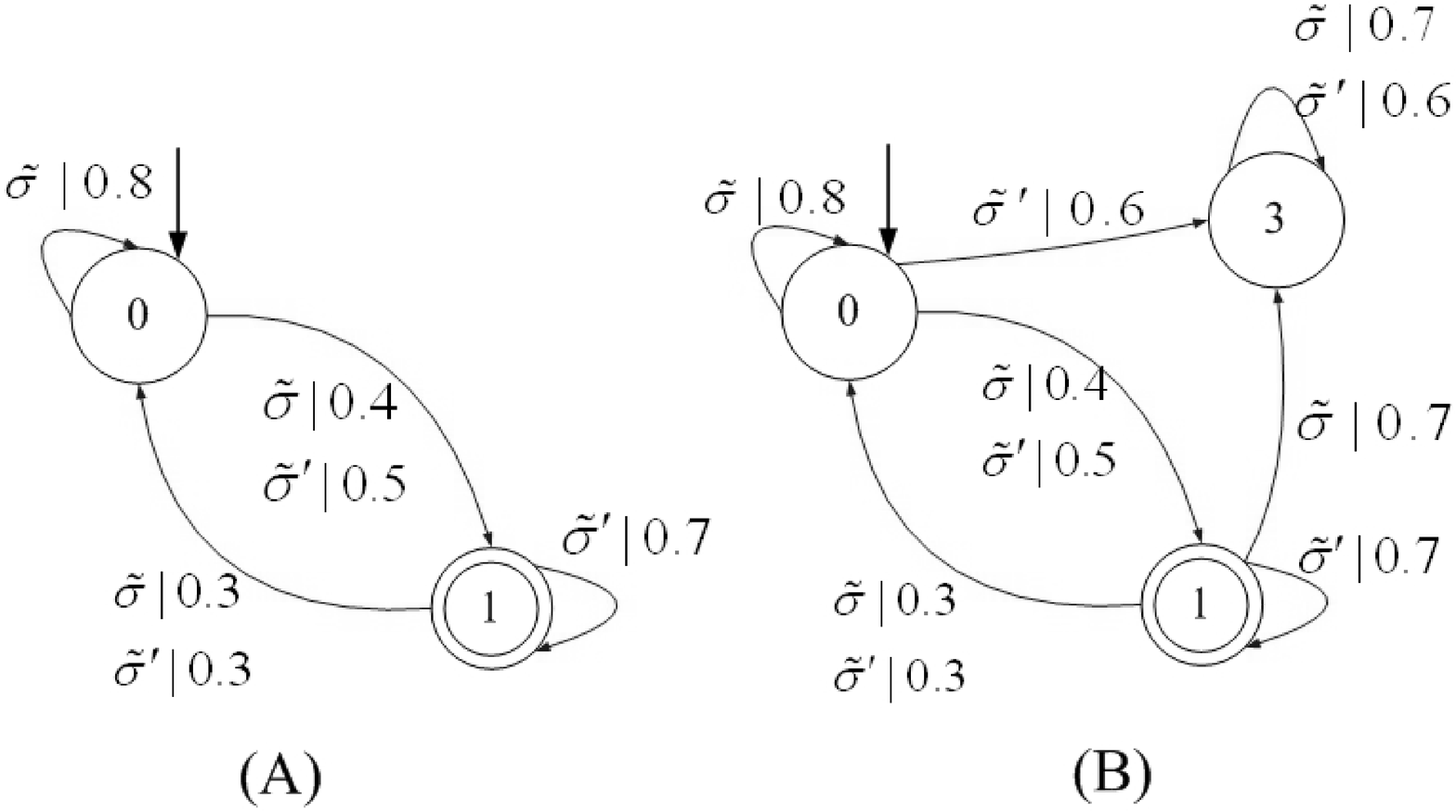}
\caption{\label{fig3} (A). FDES  $\tilde{R}$ with $\tilde{\Sigma}_{uc}(\tilde{\sigma})=0. 7$ and $\tilde{\Sigma}_{uc}(\tilde{\sigma}^{'})=0. 6$.
   (B). FDES $\tilde{R}^{+}$ constructed from $\tilde{R}$ using Algorithm 1.}
\end{figure}

\begin{lemma}
 $ \tilde{R} \subseteq \tilde{R}^{+}$.
\end{lemma}
\begin{IEEEproof}
  Suppose $|Q| = n $. Let ${\tilde{\phi}}=(I_{n*n}, \textbf{0}_{n*1})$. Then it is easy to check $\tilde{R} \subseteq_{\tilde{\phi}} \tilde{R}^{+}$.
\end{IEEEproof}
\par

\begin{lemma}
  Let $\tilde{S}$ be any fuzzy $\tilde{\Sigma}_{uc}$-compatible supervisor. Then $\tilde{R} \subseteq \tilde{S} \Rightarrow \tilde{R^{+}} \subseteq \tilde{S}$.
\end{lemma}

\begin{IEEEproof}
Suppose $\tilde{R} \subseteq_{\tilde{\phi}_{1}} \tilde{S}$ and $|Q| = n, |Y|=m$. Let
\begin{equation}
\tilde{\phi}_{2}(i)(j) =
    \begin{cases}
    \tilde{\phi}_{1}(i)(j),                                                                       & i\in[1, n], j\in[1, m],\\
    \max_{\tilde{\sigma}\in\tilde{\Sigma}}\tilde{\Sigma}_{uc}(\tilde{\sigma}),                    & i=n+1,j\in[1, m]. \\
    \end{cases}
\end{equation}
Here $\tilde{\phi}(i)(j)$ denotes the $i$th row and $j$th column element of the matrix $\tilde{\phi}$.
We show $\tilde{R^{+}} \subseteq_{\tilde{\phi}_{2}} \tilde{S}$ as follows.
\begin{enumerate}
  \item By $\tilde{R} \subseteq_{\tilde{\phi}_{1}} \tilde{S} $, we have $\tilde{q}_{0} \leq \tilde{y}_{0} \odot \tilde{\phi}_{1}^{T}$, that is,
  \[
    \tilde{q}_{0}(i) \leq \max_{j=1}^{m}\min\{\tilde{y}_{0}(j), \phi_{1}^{T}(j)(i)\}~ for  ~\forall i\in[1, n].
  \]
  Further, it is obvious that the following equation holds:
  \[
    0 \leq \max_{j=1}^{m}\min\{\tilde{y}_{0}(j), \max_{\tilde{\sigma}\in\tilde{\Sigma}}\tilde{\Sigma}_{uc}(\tilde{\sigma})\}, ~\forall i\in[1, n].
  \]
  Then by the definitions of $\tilde{\phi}_{2}$ and $\tilde{q}_{0}^{+}$, we have
  \[
    \tilde{q}_{0}^{+}(i) \leq \max_{j=1}^{m}\min\{\tilde{y}_{0}(j), \phi_{2}^{T}(j)(i) \}, ~\forall i\in[1, n+1],
  \]
  that is, $\tilde{q}_{0}^{+} \leq \tilde{y}_{0} \odot \phi_{2}^{T}$.
  \item By $\tilde{R} \subseteq_{\tilde{\phi}_{1}} \tilde{S} $, we have $\tilde{q}_{m} \odot \tilde{\phi}_{1} \leq \tilde{y}_{m}$, that is,
  \begin{equation}
  \max_{i=1}^{n}\min \{\tilde{q}_{m}(i), \tilde{\phi}_{1}(i)(j)\} \leq \tilde{y}_{m}(j), ~\forall j\in[1, m].
  \end{equation}
  Further, it is obvious that the following equation holds:
  \begin{align}
  &\max\{ \min\{0, \max_{\tilde{\sigma}\in\tilde{\Sigma}}\tilde{\Sigma}_{uc}(\tilde{\sigma})\},
          \max_{i=1}^{n}\min \{\tilde{q}_{m}(i), \tilde{\phi}_{1}(i)(j)\} \}\nonumber\\
   & =  \max_{i=1}^{n}\min \{\tilde{q}_{m}(i), \tilde{\phi}_{1}(i)(j)\}.
  \end{align}
  On the other hand, by the definitions of $\tilde{\phi}_{2}$ and $\tilde{q}_{m}^{+}$, we have
   \begin{align}
  &\max\{ \min\{0, \max_{\tilde{\sigma}\in\tilde{\Sigma}}\tilde{\Sigma}_{uc}(\tilde{\sigma})\},
         \max_{i=1}^{n}\min \{\tilde{q}_{m}(i), \tilde{\phi}_{1}(i)(j)\} \} \nonumber\\
   &   =  \max_{i=1}^{n+1}\min \{\tilde{q}_{m}^{+}(i), \tilde{\phi}_{2}(i)(j)\}.
 \end{align}

  From Equations (12), (13) and (14), we have
   \begin{equation}
  \max_{i=1}^{n+1}\min \{\tilde{q}^{+}_{m}(i), \tilde{\phi}_{2}(i)(j)\} \leq \tilde{y}_{m}(j), ~\forall j\in[1, m],
  \end{equation}
  \par
  that is,  $\tilde{q}^{+}_{m}\odot \tilde{\phi}_{2} \leq \tilde{y}_{m}$ holds.

  \item Suppose for any $ \tilde{\sigma}\in\tilde{\Sigma}$, the corresponding event matrices in $\tilde{R}$, $\tilde{R}^{+}$ and $\tilde{S}$ are denoted by $\tilde{\sigma}$, $\tilde{\sigma}^{+}$ and $\tilde{\sigma}^{s}$, respectively.
      By $\tilde{R} \subseteq_{\tilde{\phi}_{1}} \tilde{S} $, we have $\tilde{\phi}_{1}^{T} \odot \tilde{\sigma} \leq \tilde{\sigma}^{s} \odot \tilde{\phi}_{1}^{T}$ for any $ \tilde{\sigma}\in\tilde{\Sigma}$, that is,  $\forall i\in[1, n], \forall j\in[1, m]$, we have
      \begin{multline}
      \max_{i^{*}=1}^{n}\min\{\phi_{1}^{T}(j)(i^{*}), \tilde{\sigma}(i^{*})(i)\} \leq \\
       \max_{j^{*}=1}^{m}\min\{\tilde{\sigma}^{s}(j)(j^{*}), \tilde{\phi}_{1}^{T}(j^{*})(i)\}. 
    \end{multline}
      For convenience, we denote the left-hand and right-hand sides of the above inequality as $A(j)(i)$ and $B(j)(i)$, respectively.
      On the other hand, we need to show $\tilde{\phi}_{2}^{T} \odot \tilde{\sigma}^{+} \leq \tilde{\sigma}^{s} \odot \tilde{\phi}_{2}^{T}$,
      that is, $\forall i\in[1, n+1], \forall j\in[1, m]$, we need to show
      \begin{multline}
      \max_{i^{*}=1}^{n+1}\min\{\phi_{2}^{T}(j)(i^{*}), \tilde{\sigma}^{+}(i^{*})(i)\} \leq \\
      \max_{j^{*}=1}^{m}\min\{\tilde{\sigma}^{s}(j)(j^{*}), \tilde{\phi}_{2}^{T}(j^{*})(i)\}.
      \end{multline}
      \par
      For convenience, we denote the left-hand and right-hand sides of the above inequality as $C(j)(i)$ and $D(j)(i)$, respectively. Then we show $C(j)(i)\leq D(j)(i)$ by dividing into the following two cases: \par
      a)~ $i\in[1, n]$. \\
      By the definition of $\tilde{\phi}_{2}$, we have \[B(j)(i) = D(j)(i). \]
      By the definitions of $\tilde{\phi}_{2}$ and $\tilde{\sigma}^{+}$, we have
      \begin{align*}
      &C(j)(i)\\
       &= \max\{A(j)(i), \min\{\phi_{2}^{T}(j)(n+1),\tilde{\sigma}^{+}(n+1)(i)\}\} \\
      &= \max\{A(j)(i), \min\{\max_{\tilde{\sigma}\in\tilde{\Sigma}}\tilde{\Sigma}_{uc}(\tilde{\sigma}) , 0\}\} = A(j)(i). \nonumber
      \end{align*}
      Then we have $C(j)(i)\leq D(j)(i)$. \\
      b)~ $i=n+1$.\par
      By the definitions of $\tilde{\phi}_{2}$ and $\tilde{\sigma}^{+}$, we have
      \begin{align*}
      C(j)(n+1)&=\max_{i^{*}=1}^{n+1}\min\{\phi_{2}^{T}(j)(i^{*}),
            \tilde{\sigma}^{+}(i^{*})(n+1)\} \\
            &\leq \tilde{\Sigma}_{uc}(\tilde{\sigma}).
      \end{align*}
     For $\tilde{S}$ is fuzzy $\tilde{\Sigma}_{uc}$-compatible, together with the definition of $\tilde{\phi}_{2}$, we have
     \begin{align*}
     D(j)(n+1)&=  \max_{j^{*}=1}^{m}\min\{\tilde{\sigma}^{s}(j)(j^{*}),
            \tilde{\phi}_{2}^{T}(j^{*})(n+1)\} \\
            &\geq \tilde{\Sigma}_{uc}(\tilde{\sigma}).
     \end{align*}
     Then we have $C(j)(i)\leq D(j)(i)$. Therefore,  $\forall i\in[1, n+1], \forall j\in[1, m]$, $C(j)(i)\leq D(j)(i)$ holds, that is, $\tilde{\phi}_{2}^{T} \odot \tilde{\sigma}^{+} \leq \tilde{\sigma}^{s} \odot \tilde{\phi}_{2}^{T}$   holds.
\end{enumerate}
That is, we complete the proof of the lemma.
\end{IEEEproof}

The following theorem provides a necessary and sufficient condition for the existence of fuzzy supervisors.
\begin{theorem}
  Given an uncontrolled system $\tilde{G}$, and specification $\tilde{R}$, there exists a fuzzy $\tilde{\Sigma}_{uc}$-compatible supervisor $\tilde{S}$ such that $\tilde{G}||\tilde{S}\sim \tilde{R}$ if and only if $\tilde{R}\subseteq \tilde{G}$ and $\tilde{G}||\tilde{R}^{+}\subseteq \tilde{R}$, where $\tilde{R}^{+}$ has been defined in Algorithm 1.
\end{theorem}
\begin{IEEEproof}
  For sufficiency, by Lemma 3 we have $\tilde{R}\subseteq \tilde{R^{+}}$, which together with $\tilde{R}\subseteq \tilde{G}$ implies $\tilde{R}\subseteq \tilde{G}||\tilde{R}^{+}$ by Proposition 4.
  As $\tilde{G}||\tilde{R}^{+}\subseteq \tilde{R}$ holds,  we have $\tilde{G}||\tilde{R}^{+}\sim \tilde{R}$. Since $\tilde{R}^{+}$ is fuzzy $\tilde{\Sigma}_{uc}$-compatible, we can choose $\tilde{S}$ to be $\tilde{R}^{+}$. Then $\tilde{G}||\tilde{S}\sim \tilde{R}$ holds. \par
  For necessity, $\tilde{G}||\tilde{S}\sim \tilde{R}$ implies $\tilde{G}||\tilde{S}\subseteq \tilde{R}$ and $\tilde{R}\subseteq \tilde{G}||\tilde{S}$, which further implies $\tilde{R}\subseteq \tilde{G}$ and $\tilde{R}\subseteq \tilde{S}$ by Proposition 5. It remains to show $\tilde{G}||\tilde{R}^{+}\subseteq \tilde{R}$. By Lemma 4 $\tilde{R}\subseteq \tilde{S}$ implies $\tilde{R}^{+} \subseteq \tilde{S}$. By Corollary 2, $\tilde{R}^{+} \subseteq \tilde{S}$ implies $\tilde{G}||\tilde{R}^{+} \subseteq \tilde{G}||\tilde{S}$, which together with $\tilde{G}||\tilde{S}\subseteq \tilde{R}$, implies  $\tilde{G}||\tilde{R}^{+}\subseteq \tilde{R}$ by Proposition 1.
\end{IEEEproof}
\begin{remark}
  The condition of the existence of  supervisors for crisp DESs has been studied by Zhou and Kumar \cite{zhou-3}. Theorem 1 generalizes the results to FDESs. Theorem 1 shows that the problem of verifying the existence of fuzzy  supervisors can be reduced to the problem of verifying the fuzzy simulation relations, which can be solved by the search algorithm mentioned in Section  \uppercase\expandafter{\romannumeral3} (Subsection B).
 Whenever the supervisors exist, $\tilde{R}^{+}$ can serve as a supervisor.

\end{remark}
\par
 From Theorem 1, we present the following definition to characterize the achievable specifications by fuzzy simulation-equivalence control.
\begin{definition}
  Given an uncontrolled system $\tilde{G}$ with the uncontrollable set $\tilde{\Sigma}_{uc}$ and a specification $\tilde{R}$, $\tilde{R}$ is called \emph{fuzzy simulation-based controllable} with respect to $\tilde{G}$ and $\tilde{\Sigma}_{uc}$ if
   $\tilde{R}\subseteq \tilde{G}$ and $\tilde{G}||\tilde{R}^{+}\subseteq \tilde{R}$ hold.
\end{definition}
\par
The following example illustrates the fuzzy simulation-equivalence control for a specification which can be expressed by fuzzy simulation equivalence but can not by fuzzy language equivalence.
\begin{example}
     In an FDES-based disease treatment-decision support system, each of the main clinical variables of a certain disease is modeled as an FDES, in which the states denote the conditions of the clinical variable, such as ``poor", ``not bad", ``good", etc., and the events denote treatment regimens \cite{treatplaning}, \cite{treatment-2}, \cite{treatment-3}.
     Let a clinical variable be modelled as $\tilde{G}=\{\tilde{X}, \tilde{\Sigma}, \tilde{\alpha}, \tilde{x}_{0}, \tilde{x}_{m}\}$ (as shown in Fig. 4). Suppose that the first and second crisp states of $\tilde{G}$ denote ``bad"  and ``good", respectively.
    The initial state is $\tilde{x}_{0}=[1~0]$ and the marked state is $\tilde{x}_{m}=[1~1]$.  $\tilde{\Sigma}=\{\tilde{\sigma}, \tilde{\sigma}^{'}\}$, denoting the candidate treatment regimens, are fuzzy events with $\tilde{\Sigma}_{uc}(\tilde{\sigma})=0. 8$ and $\tilde{\Sigma}_{uc}(\tilde{\sigma}^{'})=0. 1$. The corresponding  matrices of the events are:
    \[
                                 \tilde{\sigma}_{1}=\left[
                                                \begin{array}{cc}
                                                   0.4 & 0.8 \\
                                                   0   & 0.4 \\
                                                \end{array}
                                             \right],
                                             \tilde{\sigma}_{1}^{'}=\left[
                                                \begin{array}{ccc}
                                                   0.4 & 0.9 \\
                                                   0.4 & 0.4\\
                                               \end{array}
                                             \right].
   \]
   \par
Besides the high cure rate, the low recurrence rate of a treatment regimen is another important desired specification in medical treatment.

 Suppose the desired treatment specification is that the recurrence rate should be no greater than $20\%$. Then the specification can be modeled as
      $\tilde{R}=\{\tilde{Q}, \tilde{\Sigma}, \tilde{\beta}, \tilde{q}_{0}, \tilde{q}_{m}\}$. Here the initial state,  the final state and the treatment regimen $\tilde{\sigma}$ are equal to those in the uncontrolled system. The corresponding treatment regimen $\tilde{\sigma}_{'}$ matrix is:
   \[
                                      \tilde{\sigma}_{2}^{'}=\left[
                                                \begin{array}{ccc}
                                                   0.4 & 0.9 \\
                                                   0.2 & 0.4 \\
                                                \end{array}
                                             \right].
\]
According to Equations (1) and (2), we can easily get the system language $L_{\tilde{G}}$ and the specification language $L_{\tilde{R}}$ as follows.
\[
     L_{\tilde{G}} = \frac{1}{\epsilon} + \frac{0.8}{\sigma} + \frac{0.9}{\sigma_{'}}  + \frac{0.4}{\tilde{s}\ ( |\tilde{s}| \geq 2 ) } = L_{\tilde{R}}.
\]
Hence, if we use fuzzy language equivalence as system behavioral equivalence, then the specification is directly achieved without control.
However, as mentioned above, the system behavior is not satisfied. Therefore, under these circumstances, the fuzzy language-equivalence control does not work and the fuzzy simulation-equivalence control is required.

\begin{figure}
\centering
\includegraphics[width=0.48\textwidth]{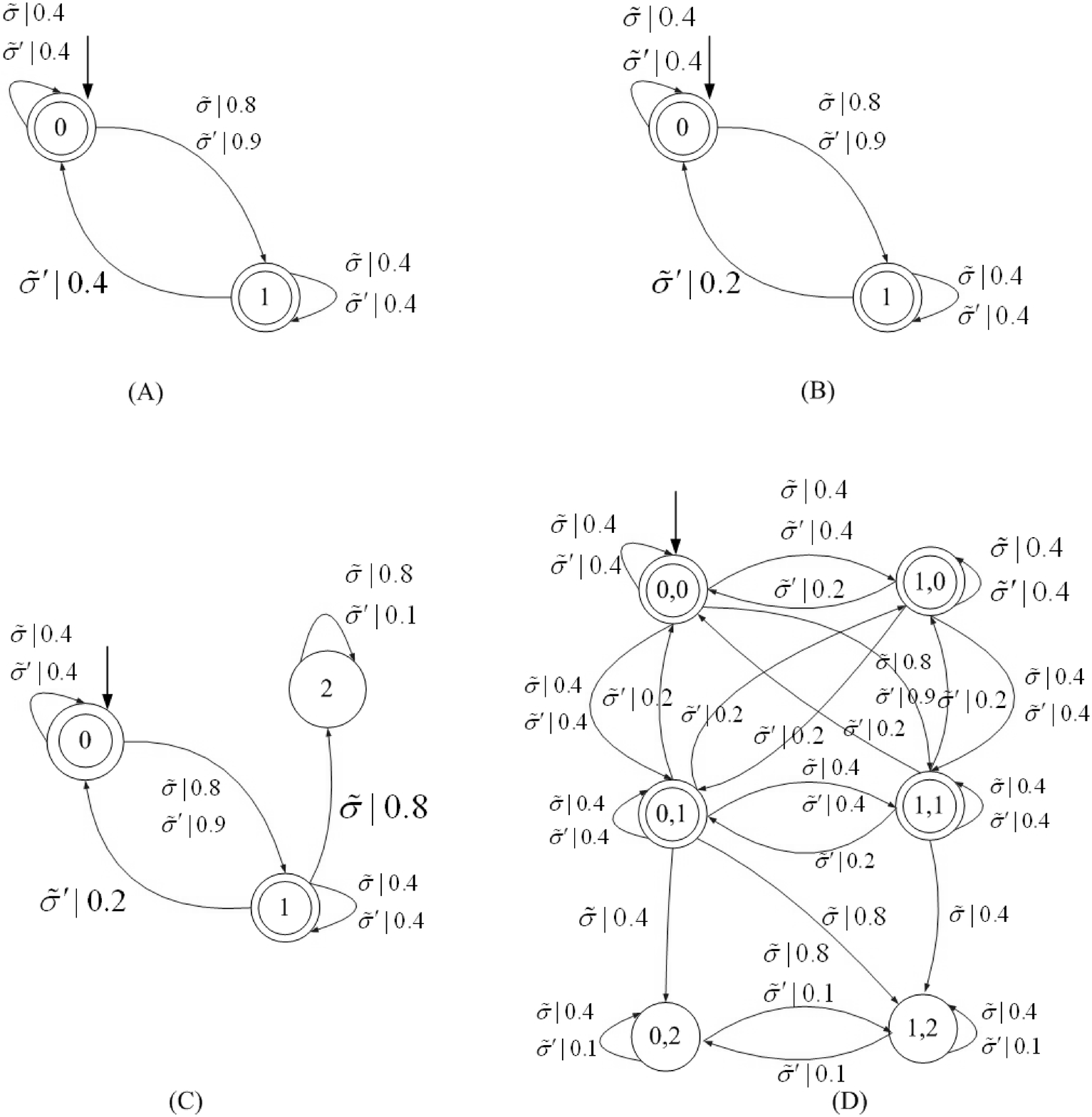}
\caption{\label{fig4} (A).The system $\tilde{G}$ with $\tilde{\Sigma}_{uc}(\tilde{\sigma})=0. 8$ and $\tilde{\Sigma}_{uc}(\tilde{\sigma}^{'})=0. 1$.  (B).The specification $\tilde{R}$. (C).$\tilde{R}^{+}$ constructed from $\tilde{R}$ using Algorithm 1
(D).$\tilde{G}||\tilde{R}^{+}$. \ \ \  Using the searching algorithm in Section \uppercase\expandafter{\romannumeral3}, we get $\tilde{R} \subseteq \tilde{G}$ and $\tilde{G}||\tilde{R}^{+} \subseteq \tilde{R}$.}
\end{figure}

Next, we consider whether the specification can be achieved by the fuzzy simulation-equivalence control or not.\par
Firstly, using the searching algorithm mentioned in Section \uppercase\expandafter{\romannumeral3} (Subsection B), we get the fuzzy simulation relation
\[
    \tilde{\phi} = \left[
                                             \begin{array}{ccc}
                                                   1 & 0 \\
                                                   0 & 1 \\
                                               \end{array}
                     \right],
\]
such that $\tilde{R} \subseteq_{\tilde{\phi}} \tilde{G}$. \par
Afterwards, we verify whether or not $\tilde{G}||\tilde{R}^{+}\subseteq \tilde{R}$ holds.
First, according to Algorithm 1, we construct the $\tilde{R}^{+}$ as follows:
   \[
       \tilde{q}_{0+}=[1~0~0]~~,~~\tilde{q}_{m+}=[1~1~0],
  \]
   \[
   \tilde{\sigma}_{+}=\left[
                                                \begin{array}{cccc}
                                                   0.4 & 0.8 & 0 \\
                                                   0   & 0.4 & 0.8 \\
                                                   0   & 0   & 0.8\\
                                                \end{array}
                                             \right]~,~
                                             \tilde{\sigma}_{+}^{'}=\left[
                                                \begin{array}{cccc}
                                                   0.4 & 0.9 & 0\\
                                                   0.2 & 0.4 & 0\\
                                                   0 & 0 & 0. 1\\
                                                \end{array}
                         \right],
\]
where $\tilde{q}_{0+}$ and $\tilde{q}_{m+}$ are the initial state and final state, respectively, and $\tilde{\sigma}_{+}$ and $\tilde{\sigma}_{+}^{'}$ are the corresponding matrices of events $\tilde{\sigma}$ and $\tilde{\sigma}^{'}$ in $\tilde{R}^{+}$. \par
Then we get $\tilde{G}||\tilde{R}^{+}$:
   \[
       \tilde{q}_{0GR}=\tilde{x}_{0} \tilde{\otimes}\tilde{q}_{0+} =[1~0~0~0~0~0],
       \]
       \[
        \tilde{q}_{mGR}= \tilde{x}_{m} \tilde{\otimes}\tilde{q}_{m+} =[1~1~0~1~1~0],
   \]
   \[
   \tilde{\sigma}_{GR}= 
                                        \left[
                                                \begin{array}{cccccc}
                                                   0.4 & 0.4 & 0  & 0.4 & 0.8 & 0 \\
                                                   0   & 0.4 &0.4 & 0   & 0.4 & 0.8 \\
                                                   0   & 0  & 0.4 & 0   & 0   & 0.8 \\
                                                   0   & 0  & 0 & 0.4   & 0.4 & 0  \\
                                                   0   & 0  & 0 & 0     & 0.4 & 0.4  \\
                                                   0   & 0  & 0 & 0     & 0  & 0.4  \\
                                                \end{array}
                                        \right],
  \]
  \[
      \tilde{\sigma}_{GR}^{'}= 
                                             \left[
                                             \begin{array}{cccccc}
                                                   0.4 & 0.4 & 0  & 0.4 & 0.9 & 0 \\
                                                   0.2 & 0.4 & 0  & 0.2 & 0.4 & 0 \\
                                                   0   & 0  & 0.1 & 0   & 0   & 0.1 \\
                                                   0.2 & 0.2& 0 & 0.4   & 0.4 & 0  \\
                                                   0.2 & 0.2& 0 & 0.2   & 0.4 & 0  \\
                                                   0   & 0 & 0.1& 0     & 0  & 0.1  \\
                                             \end{array}
                                             \right],
  \]
  where $\tilde{q}_{0GR}$ and $\tilde{q}_{mGR}$ are the initial state and final state, respectively, and $\tilde{\sigma}_{GR}$ and $\tilde{\sigma}_{GR}^{'}$ are the corresponding matrices of events $\tilde{\sigma}$ and $\tilde{\sigma}^{'}$ in $\tilde{G}||\tilde{R}^{+}$. \par
Using the searching algorithm mentioned in Section \uppercase\expandafter{\romannumeral3} (Subsection B), we get the fuzzy simulation relation
\[
    \tilde{\phi} = \left[
                                             \begin{array}{cccccc}
                                                   1   & 0.4 & 0.4 & 0.4 & 0.4   & 0.4 \\
                                                   0.4   & 0.4 & 0.4 & 0.4 & 0.9   & 0.4 \\
                                               \end{array}
                     \right]^{T},
\]
such that $\tilde{G}||\tilde{R}^{+} \subseteq_{\tilde{\phi}} \tilde{R}$.
Therefore, $\tilde{R}$ is simulation-based controllable, and $\tilde{R}^{+}$ serves as the supervisor $\tilde{S}$ to ensure that
$\tilde{G}||\tilde{S} \sim \tilde{R}$. That is, the  specification can be achieved by fuzzy simulation-equivalence control.
\end{example}

\par
  We have discussed the ``target" control problem, which aims to ensure $\tilde{G}||\tilde{S}\sim \tilde{R}$, or equivalently $\tilde{R}\subseteq\tilde{G}||\tilde{S}\subseteq \tilde{R}$. We continue to consider a more general ``range" control problem, which aims to ensure $\tilde{R}_{1}\subseteq\tilde{G}||\tilde{S}\subseteq \tilde{R}_{2}$, where automaton $\tilde{R}_{1}$ and  automaton $\tilde{R}_{2}$ specify the minimally and maximally desired system behavior, respectively.  $\tilde{R}_{1}=\tilde{R}_{2}=\tilde{R}$ holds in the ``target" control problem. The following theorem discusses the ``range" control problem and presents a necessary and sufficient condition for the existence of the ``range" supervisor.
\begin{theorem}
    Given an uncontrolled system $\tilde{G}$, and the lower bond specification $\tilde{R}_{1}$ and upper bound specification $\tilde{R}_{2}$ such that $\tilde{R}_{1}\subseteq \tilde{R}_{2}$, there exists a fuzzy $\tilde{\Sigma}_{uc}$-compatible supervisor $\tilde{S}$ such that $\tilde{R}_{1} \subseteq\tilde{G}||\tilde{S}\subseteq\tilde{R}_{2}$ if and only if $\tilde{R}_{1}\subseteq \tilde{G}$ and $\tilde{G}||\tilde{R}_{1}^{+}\subseteq \tilde{R}_{2}$.
\end{theorem}
\begin{IEEEproof}
  For sufficiency, by Lemma 3 we have $\tilde{R}_{1}\subseteq \tilde{R}_{1}^{+}$. Together with $\tilde{R}_{1}\subseteq \tilde{G}$, it implies $\tilde{R}_{1}\subseteq \tilde{G}||\tilde{R}_{1}^{+}$ by Proposition 4. As $\tilde{G}||\tilde{R}_{1}^{+}\subseteq \tilde{R}_{2}$ holds, we have $\tilde{R}_{1}\subseteq \tilde{G}||\tilde{R}_{1}^{+} \subseteq \tilde{R}_{2}$.
   Since $\tilde{R}_{1}^{+}$ is $\tilde{\Sigma}_{uc}$-compatible, let $\tilde{S}$ be $\tilde{R}_{1}^{+}$, and we have $\tilde{R}_{1} \subseteq\tilde{G}||\tilde{S}\subseteq\tilde{R}_{2}$. \par
  For necessity, by Proposition 5, $\tilde{R}_{1} \subseteq\tilde{G}||\tilde{S}$ implies $\tilde{R}_{1}\subseteq \tilde{G}$ and $\tilde{R}_{1}\subseteq \tilde{S}$. It remains to show $\tilde{G}||\tilde{R}_{1}^{+}\subseteq \tilde{R}_{2}$. By Lemma 4, $\tilde{R}_{1}\subseteq \tilde{S}$ implies $\tilde{R}_{1}^{+}\subseteq \tilde{S}$. By Corollary 2, $\tilde{R}_{1}^{+}\subseteq \tilde{S}$ implies $\tilde{G}||\tilde{R}_{1}^{+}\subseteq \tilde{G}||\tilde{S}$, which together with $\tilde{G}||\tilde{S}\subseteq \tilde{R}_{2}$, implies $\tilde{G}||\tilde{R}_{1}^{+}\subseteq \tilde{R}_{2}$ by Proposition 1.
\end{IEEEproof}

\par
The following example illustrates the above results. For convenience to calculate by hand, the following example is simplified by restricting all the elements in state vectors and event matrices to 0 or 1.
\begin{example}
 Consider an uncontrolled system $\tilde{G}$ with the minimally behavior $\tilde{R}_{1}$ and maximally behavior $\tilde{R}_{2}$ and the uncontrollable set $\tilde{\Sigma}_{uc}({\tilde{\sigma}_{1}}) = \tilde{\Sigma}_{uc}({\tilde{\sigma}_{2}}) = 0, \tilde{\Sigma}_{uc}({\tilde{\sigma}_{3}}) = 1$.
  Due to the limited space, we do not present the state vectors and event matrices of $\tilde{G}$, $\tilde{R}_{1}$ and $\tilde{R}_{2}$. For the detail, see  Fig. 5-(A-C).\par
   It is obvious that $\tilde{R}_{1}\subseteq \tilde{G}$ holds. We need to verify whether $\tilde{G}||\tilde{R}_{1}^{+}\subseteq \tilde{R}_{2}$ holds or not.
  First, following  Algorithm 1, we construct $\tilde{R}^{+}$, as shown in Fig. 5-(D).  Then, we further obtain $\tilde{G}||\tilde{R}^{+}$, as shown in Fig. 5-(E). Using the searching algorithm in Section \uppercase\expandafter{\romannumeral3}, we obtain that
  $\tilde{G}||\tilde{R}_{1}^{+}\subseteq \tilde{R}_{2}$ does not hold. Thus, the ``range" control problem of $\tilde{G}$ has no solution. \par
  If the uncontrollable set is revised to $\tilde{\Sigma}_{uc}({\tilde{\sigma}_{1}}) = \tilde{\Sigma}_{uc}({\tilde{\sigma}_{3}}) = 0$ and $
  \tilde{\Sigma}_{uc}({\tilde{\sigma}_{2}}) = 1$, following the aforementioned  steps, we obtain that both $\tilde{R}_{1}\subseteq \tilde{G}$ and $\tilde{G}||\tilde{R}_{1}^{+}\subseteq \tilde{R}_{2}$ hold. Therefore, the ``range" control problem of $\tilde{G}$ has at least  one solution $R_{1}^{+}$.

\begin{figure}
\centering
\includegraphics[width=0.4\textwidth]{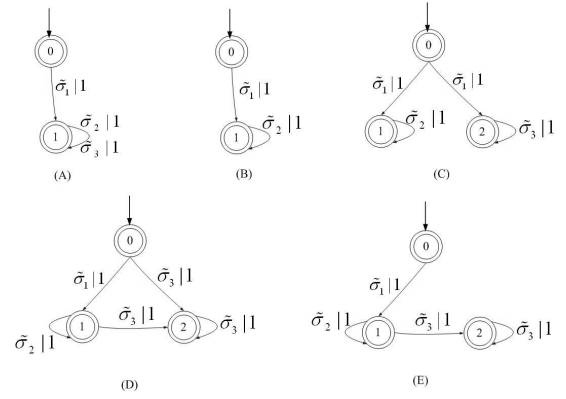}
\caption{\label{fig5} (A).The uncontrolled system $\tilde{G}$  (B).The the minimally behavior $\tilde{R}_{1}$ (C). maximally behavior $\tilde{R}_{2}$ (D) $\tilde{R}_{1}^{+}$ constructed from $\tilde{R}_{1}$ using Algorithm 1
(E).$\tilde{G}||\tilde{R}_{1}^{+}$.}
\end{figure}

\end{example}

\subsection{Fuzzy Language-equivalence Control and Fuzzy Simulation -equivalence Control}
In this subsection, we continue to investigate the relations between fuzzy language-equivalence control and fuzzy simulation-equivalence control. \par

    Fuzzy language-equivalence control  has been studied by Qiu \cite{Qiu-1}, \cite{Qiu-3} in detail, the objective of which is ensure that  the controlled system $L_{\tilde{S}/\tilde{G}}$ is fuzzy language equivalent with the given specification $pr(\tilde{K})$.

Qiu \cite{Qiu-1} presented the following notion to characterize  achievable languages by fuzzy language-equivalence control.
\par
\begin{definition}
  Let $\tilde{K}$ and $\tilde{M}$ be fuzzy languages over fuzzy event set $\tilde{\Sigma}$ and $pr(\tilde{M})=\tilde{M}$. $\tilde{K}$ is said to be \emph{fuzzy language-based controllable} with respect to $\tilde{M}$ and $\tilde{\Sigma}_{uc}$ if for any $\tilde{s}\in\tilde{\Sigma}^{*}$ and for any $\tilde{\sigma}\in\tilde{\Sigma}$, the following equation holds:
  \begin{equation}
  \min\{ pr(\tilde{K})(\tilde{s}), \tilde{\Sigma}_{uc}(\tilde{\sigma}), \tilde{M}(\tilde{s}\tilde{\sigma})  \} \leq pr(\tilde{K})(\tilde{s}\tilde{\sigma}).
  \end{equation}
   Equation (18) is called \emph{fuzzy controllability condition} of $\tilde{K}$ with respect to $\tilde{M}$ and $\tilde{\Sigma}_{uc}$ in \cite{Qiu-1}. In order to emphasize the difference of the fuzzy language-equivalence control studied in \cite{Qiu-1} and the fuzzy simulation-equivalence control studied in this paper, henceforth, we called  Equation (18) as \emph{fuzzy language-based controllability condition} of $\tilde{K}$ with respect to $\tilde{M}$ and $\tilde{\Sigma}_{uc}$.
\end{definition}
\par

Before giving the main theorem, we present two useful lemmas.
\begin{lemma}
  Given a fuzzy $\tilde{\Sigma}_{uc}$-compatible supervisor $\tilde{S}$, then $\forall \tilde{s}\in\tilde{\Sigma}^{*}$ and $\forall \tilde{\sigma}\in \tilde{\Sigma}$, $L(\tilde{S})(\tilde{s}\tilde{\sigma}) \geq \min\{\tilde{\Sigma}_{uc}(\tilde{\sigma}), $ $L(\tilde{S})(\tilde{s})\}$.
\end{lemma}
\begin{IEEEproof}
  Suppose that after the occurrence of the fuzzy event string $\tilde{s}$, the automaton $\tilde{S}$ turns to the fuzzy state $\tilde{q}=[q_{0}, q_{1}, \ldots, q_{n}]$. Then by the definition of fuzzy language, we have $\max_{i=1}^{n}(q_{i}) = L(\tilde{S})(\tilde{s})$. Without loss of generality,
  assume $q_{i^{*}} = L(\tilde{S})(\tilde{s})$. Since $\tilde{S}$ is $\tilde{\Sigma}_{uc}$-compatible supervisor, $\max_{j=1}^{n}\tilde{\sigma}(i^{*})(j) \geq \tilde{\Sigma}_{uc}(\tilde{\sigma})$. Without loss of generality, assume $\tilde{\sigma}(i^{*})(j^{*}) \geq \tilde{\Sigma}_{uc}(\tilde{\sigma})$. Then we have
  \begin{align}
    L(\tilde{S})(\tilde{s}\tilde{\sigma}) &= \max_{i^{'}=1}^{n}\{ \max_{j^{'}=1}^{n}\min\{q_{j^{'}}, \sigma(j^{'})(i^{'})\}\} \nonumber \\
    &\geq \min\{q_{i^{*}}, \sigma(i^{*})(j^{*})\} \nonumber \\
    & \geq \min\{\tilde{\Sigma}_{uc}(\tilde{\sigma}), L(\tilde{S})(\tilde{s}) \}. \nonumber
  \end{align}
\end{IEEEproof}
\begin{lemma}
     $ \tilde{G}_{1}\subseteq \tilde{G}_{2}\Rightarrow L(\tilde{G}_{1})\leq L(\tilde{G}_{2})$ \ ;\
     $\ \ \ \tilde{G}_{1}\sim \tilde{G}_{2}\Rightarrow L(\tilde{G}_{1})=L(\tilde{G}_{2})$.
\end{lemma}
\begin{IEEEproof}
We can  refer to Theorem 5. 3 in \cite{bi_fuzzy_automata}.
\end{IEEEproof}

\begin{theorem}
  Given fuzzy automata $\tilde{G}$ and $\tilde{R}$ with $\tilde{\Sigma}_{uc}$, if $\tilde{R}$ is fuzzy simulation-based controllable with respect to $\tilde{G}$ and $\tilde{\Sigma}_{uc}$, then $L(\tilde{R})$ is fuzzy language-based controllable with respect to $L(\tilde{G})$ and $\tilde{\Sigma}_{uc}$.
\end{theorem}
\begin{IEEEproof}
  Since $\tilde{R}$ is fuzzy simulation-based controllable, we assume there exists a fuzzy $\tilde{\Sigma}_{uc}$-compatible supervisor $\tilde{S}$ such that  $\tilde{G}||\tilde{S} \sim \tilde{R}$. \par
  By Lemma 5, for any $\tilde{s}\in\tilde{\Sigma}^{*}$ and $\forall \tilde{\sigma}\in \tilde{\Sigma}$, we have
  \[
   L(\tilde{S})(\tilde{s}\tilde{\sigma}) \geq \min\{\tilde{\Sigma}_{uc}(\tilde{\sigma}), L(\tilde{S})(\tilde{s})\},
  \]
  which implies
  \begin{multline}
    \min\{L(\tilde{G})(\tilde{s}\tilde{\sigma}), L(\tilde{S})(\tilde{s}\tilde{\sigma})\} \geq \\
     \min\{L(\tilde{G})(\tilde{s}\tilde{\sigma}), \tilde{\Sigma}_{uc}(\tilde{\sigma}), L(\tilde{S})(\tilde{s})\}.
  \end{multline}
  By Proposition 3 and Lemma 6, we have
  \begin{align*}
 & \min\{L(\tilde{G})(\tilde{s}\tilde{\sigma}), L(\tilde{S})(\tilde{s}\tilde{\sigma})\}\\
 & = L(\tilde{G})\tilde{\cap}L(\tilde{S}) (\tilde{s}\tilde{\sigma})
   =  L(\tilde{G}||\tilde{S})(\tilde{s}\tilde{\sigma}) =
  L(\tilde{R})(\tilde{s}\tilde{\sigma}),
  \end{align*}
  which further implies
  $L(\tilde{S})\geq L(\tilde{R})$, that is, $L(\tilde{S})(\tilde{s}) \geq L(\tilde{R})(\tilde{s})$.
  Then it is obvious that the following equation holds:
  \begin{multline}
   \min\{L(\tilde{G})(\tilde{s}\tilde{\sigma}), \tilde{\Sigma}_{uc}(\tilde{\sigma}), L(\tilde{S})(\tilde{s})\} \geq \\
    \min\{L(\tilde{G})(\tilde{s}\tilde{\sigma}), \tilde{\Sigma}_{uc}(\tilde{\sigma}), L(\tilde{R})(\tilde{s})\}.
  \end{multline}
  Therefore, with Equations (19) and (20), we have
  \[
  L(\tilde{R})(\tilde{s}\tilde{\sigma}) \geq \min\{L(\tilde{G})(\tilde{s}\tilde{\sigma}), \tilde{\Sigma}_{uc}(\tilde{\sigma}), L(\tilde{R})(\tilde{s})\}.
  \]
  That is, $L(\tilde{R})$ is fuzzy language-based controllable with respect to $L(\tilde{G})$ and $\tilde{\Sigma}_{uc}$.
\end{IEEEproof}
\begin{remark}
  Theorem 3 characterizes the relation between fuzzy language-equivalence controllability and fuzzy simulation-equivalence controllability.
    In the fuzzy simulation-equivalence control, the specification is given by a fuzzy automaton $\tilde{R}$, whereas, in the fuzzy language-equivalence control, the specification is given by a fuzzy language $\tilde{K}$.
    If the specification $\tilde{R}$ is achievable by  the fuzzy simulation-equivalence control, then $pr(\tilde{K}) = L(\tilde{R})$ is achievable by the fuzzy language-equivalence control.
    However, the inverse does not hold. So in this sense, we can say the fuzzy simulation-equivalence control is more precise than the fuzzy language-equivalence control.
\end{remark}
\par
The rest of this section gives a counter-example to illustrate further that the fuzzy language-based controllability does not imply the corresponding fuzzy simulation-based controllability.
\begin{example}
   Let the uncontrolled system $\tilde{G}=\{\tilde{X}, \tilde{\Sigma}, \tilde{\alpha},$ $ \tilde{x}_{0}, \tilde{x}_{m}\}$ and the specification
   $\tilde{R}=\{\tilde{Q}, \tilde{\Sigma}, \tilde{\beta}, \tilde{q}_{0}, \tilde{q}_{m}\}$, where $\tilde{\Sigma}=\{\tilde{\sigma}, \tilde{\sigma}^{'}\}$, $\tilde{\Sigma}_{uc}(\tilde{\sigma})=0. 8$, $\tilde{\Sigma}_{uc}(\tilde{\sigma}^{'})=0. 2$ and
   \[
   \tilde{x}_{0}=[0. 4~0. 7~0], \tilde{x}_{m}=[1~1~1],
   \]
   \[\tilde{\sigma}_{1}=\left[
                                                \begin{array}{ccc}
                                                   0 & 0 & 1 \\
                                                   0 & 0 & 1 \\
                                                   0 & 0 & 0 \\
                                                \end{array}
                                             \right],
                                             \tilde{\sigma}_{1}^{'}=\left[
                                                \begin{array}{ccc}
                                                   0 & 0 & 0 \\
                                                   0 & 0 & 1 \\
                                                   0 & 0 & 0 \\
                                               \end{array}
                                             \right],
   \]
   \[
   \tilde{q}_{0}=[0. 7~0. 7~0], \tilde{q}_{m}=[1~1~1],
   \]
   \[
   \tilde{\sigma}_{2}=\left[
                                                \begin{array}{ccc}
                                                   0 & 0 & 1 \\
                                                   0 & 0 & 0 \\
                                                   0 & 0 & 0 \\
                                                \end{array}
                                             \right],
                                             \tilde{\sigma}_{2}^{'}=\left[
                                                \begin{array}{ccc}
                                                   0 & 0 & 0 \\
                                                   0 & 0 & 1 \\
                                                   0 & 0 & 0 \\
                                                \end{array}
                                             \right],
\]
where $\tilde{\sigma}_{1}, \tilde{\sigma}_{2}$ and $\tilde{\sigma}_{1}^{'}, \tilde{\sigma}_{2}^{'}$ are the corresponding matrices of events $\tilde{\sigma}$ and $\tilde{\sigma}^{'}$  in $\tilde{G}$ and $\tilde{R}$, respectively. \par
Following the above settings, we get $L(\tilde{G})(\tilde{\epsilon})=L(\tilde{R})(\tilde{\epsilon})=0. 7$, $L(\tilde{G})(\tilde{\sigma})=L(\tilde{R})(\tilde{\sigma})=0. 7$,
$L(\tilde{G})(\tilde{\sigma^{'}})=L(\tilde{R})(\tilde{\sigma}^{'})=0. 7$ and for $\forall
\tilde{s}\in\tilde{\Sigma}^{*}$, $|\tilde{s}|\geq 2 $, $L(\tilde{G})(\tilde{s})=L(\tilde{R})(\tilde{s})=0$, that is, $L(\tilde{G})=L(\tilde{R})$. By the fuzzy language-based controllability condition (Equation (18)), we get $L(\tilde{R})$ is fuzzy language-based controllable. \par
Next we show $\tilde{R}$ is not fuzzy simulation-based controllable. First, we construct the $\tilde{R}^{+}$ as follows:
   \[
       \tilde{q}_{0+}=[0. 7~0. 7~0~0], \tilde{q}_{m+}=[1~1~1~0],
   \]
   \[
   \tilde{\sigma}_{+}=\left[
                                                \begin{array}{cccc}
                                                   0 & 0 & 1 & 0 \\
                                                   0 & 0 & 0 & 0. 8 \\
                                                   0 & 0 & 0 & 0. 8\\
                                                   0 & 0 & 0 & 0. 8\\
                                                \end{array}
                                             \right],
                                             \tilde{\sigma}_{+}^{'}=\left[
                                                \begin{array}{cccc}
                                                   0 & 0 & 0 & 0. 2\\
                                                   0 & 0 & 1 & 0\\
                                                   0 & 0 & 0 & 0. 2\\
                                                   0 & 0 & 0 & 0. 2\\
                                                \end{array}
                                             \right],
\]
where $\tilde{q}_{0+}$ and $\tilde{q}_{m+}$ are the initial state and final state, respectively, and $\tilde{\sigma}_{+}$ and $\tilde{\sigma}_{+}^{'}$ are the corresponding matrices of events $\tilde{\sigma}$ and $\tilde{\sigma}^{'}$ in $\tilde{R}^{+}$.
Then we get $\tilde{G}||\tilde{R}^{+}$:
   \[
       \tilde{q}_{0GR}=[0. 4~0. 4~0~0~0. 7~0. 7~0~0~0~0~0~0] \ ,
   \]
   \[
   \ \tilde{q}_{mGR}=[1~1~1~0~1~1~1~0~1~1~1~0],
   \]
   \[
   \tilde{\sigma}_{GR}=
                                        \left[
                                                \begin{array}{ccc}
                                                    \textbf{0}_{4*4} & \textbf{0}_{4*4} & \tilde{\sigma}_{+} \\
                                                    \textbf{0}_{4*4} & \textbf{0}_{4*4} & \tilde{\sigma}_{+} \\
                                                    \textbf{0}_{4*4} & \textbf{0}_{4*4} & \textbf{0}_{4*4}\\
                                                \end{array}
                                        \right] \ ,
                                        \]
                                        \[
                                             \tilde{\sigma}_{GR}^{'}=
                                             \left[
                                             \begin{array}{cccc}
                                                   \textbf{0}_{4*4} & \textbf{0}_{4*4} & \textbf{0}_{4*4} \\
                                                   \textbf{0}_{4*4} & \textbf{0}_{4*4} & \tilde{\sigma}_{+}^{'} \\
                                                   \textbf{0}_{4*4} & \textbf{0}_{4*4} & \textbf{0}_{4*4}\\

                                             \end{array}
                                             \right],
  \]
where $\tilde{q}_{0GR}$ and $\tilde{q}_{mGR}$ are the initial state and final state, respectively, and $\tilde{\sigma}_{GR}$ and $\tilde{\sigma}_{GR}^{'}$ are the corresponding matrices of events $\tilde{\sigma}$ and $\tilde{\sigma}^{'}$ in $\tilde{G}||\tilde{R}^{+}$. \par
Suppose there exists a fuzzy relation $\tilde{\phi}=[\phi_{1}^{T}, \phi_{2}^{T}, \phi_{3}^{T}]$, where $\phi_{i}, i\in\{1, 2, 3\}$ is a row vector of order 12, such that $\tilde{G}||\tilde{R}^{+}\subseteq_{\tilde{\phi}} \tilde{R}$, that is, the following equations hold.
\begin{equation}
\left[
    \begin{array}{c}
    \phi_{1}\\
    \phi_{2}\\
    \phi_{3}\\
    \end{array}
\right]
\odot \tilde{\sigma}_{GR} \leq \tilde{\sigma}_{2} \odot
\left[
    \begin{array}{c}
    \phi_{1}\\
    \phi_{2}\\
    \phi_{3}\\
    \end{array}
\right] =
\left[
    \begin{array}{c}
    \phi_{3}\\
    0\\
    0\\
    \end{array}
\right],
\end{equation}

\begin{equation}
\left[
    \begin{array}{c}
    \phi_{1}\\
    \phi_{2}\\
    \phi_{3}\\
    \end{array}
\right]
\odot \tilde{\sigma}_{GR}^{'} \leq \tilde{\sigma}_{2}^{'} \odot
\left[
    \begin{array}{c}
    \phi_{1}\\
    \phi_{2}\\
    \phi_{3}\\
    \end{array}
\right] =
\left[
    \begin{array}{c}
    0\\
    \phi_{3}\\
    0\\
    \end{array}
\right],
\end{equation}

\begin{multline}
[0. 4~0. 4~0~0~0. 7~0. 7~0~0~0~0~0~0]= \tilde{q}_{0GR} \leq \\
\tilde{q}_{0} \odot
\left[
    \begin{array}{c}
    \phi_{1}\\
    \phi_{2}\\
    \phi_{3}\\
    \end{array}
\right]
= [0. 7~0. 7~0]\odot
\left[
    \begin{array}{c}
    \phi_{1}\\
    \phi_{2}\\
    \phi_{3}\\
    \end{array}
\right].
\end{multline}
From Equation (21) and Equation (22), we get $\phi_{2}(5)=0$ and $\phi_{1}(5)=0$~~($\phi_{1}(5)$ and $\phi_{2}(5)$ denote the $5$th entry of $\phi_{1}$ and $\phi_{2}$, respectively),
which contradict with Equation (23). Hence, $\tilde{G}||\tilde{R}^{+}\subseteq \tilde{R}$ does not hold, that
is, $\tilde{R}$ is not simulation-based controllable.
\end{example}

\section{Conclusion}
 FDESs were first proposed by Lin and Ying \cite{Feng-1}, and since then FDESs have been well investigated by many authors (for instance, \cite{Cao07},  \cite{LQ09}, \cite{Wang10}, \cite{Luo12}, \cite{EK12}). The supervisory control theory of FDESs for fuzzy language equivalence was developed by Qiu \cite{Qiu-1} as well as Cao and Ying \cite{Cao-1}, respectively. As the fuzzy language equivalence has  limited expressiveness, in this paper we have established the supervisory control theory of FDESs for fuzzy simulation equivalence whose expressiveness is stronger than that of fuzzy language equivalence.
More specifically, the fuzzy simulation and fuzzy simulation equivalence of FDESs have been formulated. Several basic properties of fuzzy simulation relations have been discussed. Then, we have presented a necessary and sufficient
condition for the existence of fuzzy supervisors for FDESs, and given an efficient algorithm for constructing a supervisor whenever it exists. Moreover, we have investigated the relations of the fuzzy language-based controllability and fuzzy simulation-based controllability, and the results suggest that fuzzy simulation-equivalence control is more precise than fuzzy language-equivalence control. In addition, several examples have been used to support the findings in this paper.

\par
  Since we have assumed all the events are observable by the fuzzy supervisors, a further issue worthy of consideration is to deal with fuzzy simulation-equivalence control problem under partial observation. Furthermore, dealing with the decentralized supervisory control problem of FDESs for simulation equivalence is another challenge. These problems should be also worthy of consideration in subsequent work.\par
\appendices
\section{Proof of Proposition 4}
Suppose $\tilde{G}_{3}\subseteq_{\tilde{\phi}_{1}} \tilde{G}_{1}$, $\tilde{G}_{3}\subseteq_{\tilde{\phi}_{2}} \tilde{G}_{2}$, $|X_{1}| = m$, $|X_{2}| = n$, $|X_{3}| = k$. Let $\tilde{\phi}(p)((q-1)*n+r)=\min(\tilde{\phi}_{1}(p)(q), \tilde{\phi}_{2}(p)(r) )$ for $\forall q\in[1, m], r\in[1, n], p\in[1, k]$.
Here the $\tilde{\phi}(i)(j)$ denotes the $i$th row and $j$th column element of the matrix $\tilde{\phi}$.
 We show $ \tilde{G}_{3} \subseteq_{\tilde{\phi}} \tilde{G}_{1}||\tilde{G}_{2} $ as follows.
\par
\begin{enumerate}
\item We first show $\tilde{x}_{03} \leq \tilde{x}_{01} \tilde{\otimes} x_{02} \odot \tilde{\phi^{T}}$. That is,
\begin{multline}
  \tilde{x}_{03}(p) \leq \max_{q\in[1, m]}^{r\in[1, n]}\min\{ \min\{\tilde{x}_{01}(q), \tilde{x}_{02}(r)\},\\
  \tilde{\phi^{T}}((q-1)*n+r)(p)\}, ~ \forall p\in[1, k].
\end{multline}
By the definition of $\tilde{\phi}$, we get another form for Equation (24):
\begin{multline}
  \tilde{x}_{03}(p) \leq \max_{q\in[1, m]}^{r\in[1, n]}\min\{ \tilde{x}_{01}(q), \tilde{x}_{02}(r),\\
  \tilde{\phi_{1}^{T}}(q)(p), \tilde{\phi_{2}^{T}}(r)(p)  \}, ~ \forall p\in[1, k].
\end{multline}
By $\tilde{G}_{3}\subseteq_{\tilde{\phi}_{1}} \tilde{G}_{1}$ and $\tilde{G}_{3}\subseteq_{\tilde{\phi}_{2}} \tilde{G}_{2}$, $\forall p\in[1, k]$, we have:
\begin{gather}
  \tilde{x}_{03}(p) \leq \max_{q\in[1, m]}\min\{ \tilde{x}_{01}(q), \tilde{\phi_{1}^{T}}(q)(p)\};  \\
  \tilde{x}_{03}(p) \leq \max_{r\in[1, n]}\min\{ \tilde{x}_{02}(r), \tilde{\phi_{2}^{T}}(r)(p)\}.
\end{gather}
Suppose when $q=q^{*}$ and $r=r^{*}$, the right-hand side of Equations (26) and (27) gets the maxima. Then we get
\begin{multline}
  \tilde{x}_{03}(p) \leq
  \\
   \min\{ \tilde{x}_{01}(q^{*}), \tilde{x}_{02}(r^{*}),
  \tilde{\phi_{1}^{T}}(q^{*})(p), \tilde{\phi_{2}^{T}}(r^{*})(p)\}, \nonumber
\end{multline}
which implies Equation (25). That is to say, $\tilde{x}_{03} \leq \tilde{x}_{01} \tilde{\otimes} x_{02} \odot \tilde{\phi^{T}}$ holds.

\item We continue to show $\tilde{x}_{m3} \odot \tilde{\phi} \leq \tilde{x}_{m1} \tilde{\otimes} \tilde{x}_{m2}$. That is,  $\forall q\in[1, m]$ and $\forall r\in[1, n]$,
 \begin{multline}
  \max_{p\in[1, k]}\min \{ \tilde{x}_{m3}(p), \tilde{\phi}(p)((q-1)*n+r)\} \leq \\
  \min(\tilde{x}_{m1}(q), \tilde{x}_{m2}(r)).
\end{multline}
By the definition of $\tilde{\phi}$, we get another form for  Equation (28):
 \begin{multline}
  \max_{p\in[1, k]}\min \{ \tilde{x}_{m3}(p), \tilde{\phi_{1}}(p)(q), \tilde{\phi_{2}}(p)(r) \} \leq \\
  \min(\tilde{x}_{m1}(q), \tilde{x}_{m2}(r)).
\end{multline}
By $\tilde{G}_{3}\subseteq_{\tilde{\phi}_{1}} \tilde{G}_{1}$ and $\tilde{G}_{3}\subseteq_{\tilde{\phi}_{2}} \tilde{G}_{2}$, we have
\begin{gather}
  \max_{p\in[1, k]}\min \{ \tilde{x}_{m3}(p), \tilde{\phi_{1}}(p)(q)\} \leq \tilde{x}_{m1}(q),~\forall q\in[1, m]; \nonumber \\
  \max_{p\in[1, k]}\min \{ \tilde{x}_{m3}(p), \tilde{\phi_{2}}(p)(r)\} \leq \tilde{x}_{m2}(r),~\forall r\in[1, r]. \nonumber
\end{gather}
Furthermore we get the following two equations: $~\forall q\in[1, m]$ and $ ~\forall r\in[1, r] $ :
\[
  \max_{p\in[1, k]}\min \{ \tilde{x}_{m3}(p), \tilde{\phi_{1}}(p)(q), \tilde{\phi_{2}}(p)(r)\} \leq \tilde{x}_{m1}(q);
\]
\[
  \max_{p\in[1, k]}\min \{ \tilde{x}_{m3}(p), \tilde{\phi_{1}}(p)(q), \tilde{\phi_{2}}(p)(r)\} \leq  \tilde{x}_{m2}(r),
\]
which both imply Equation (29). That is, $\tilde{x}_{m3} \odot \tilde{\phi} \leq \tilde{x}_{m1} \tilde{\otimes} \tilde{x}_{m2}$ holds.

\item Finally, we show $\tilde{\phi}^{T} \odot \tilde{\sigma}_{3} \leq \tilde{\sigma}_{1} \tilde{\otimes} \tilde{\sigma}_{2} \odot \tilde{\phi}^{T}$. That is, $\forall q\in[1, m], r\in[1, n], p\in[1, k]$,
 \begin{multline}
  \max_{p^{*}\in[1, k]}\min \{ \tilde{\phi^{T}}((q-1)*n+r)(p^{*}), \tilde{\sigma}_{3}(p^{*})(p)\} \leq \\
  \max_{q^{*}\in[1, m]}^{r^{*}\in[1, n]}\min\{ \min(\tilde{\sigma}_{1}(q)(q^{*}), \tilde{\sigma}_{2}(r)(r^{*})), \\
  \tilde{\phi^{T}}((q^{*}-1)*n+r^{*})(p) \}.
\end{multline}
By the definition of $\tilde{\phi}$, we get a simple form for Equation (30) as follows:
 \begin{multline}
  \max_{p^{*}\in[1, k]}\min \{ \tilde{\phi_{1}^{T}}(q)(p^{*}), \tilde{\phi_{2}^{T}}(r)(p^{*}), \tilde{\sigma}_{3}(p^{*})(p)\} \leq \\
  \max_{q^{*}\in[1, m]}^{r^{*}\in[1, n]}\min\{\tilde{\sigma}_{1}(q)(q^{*}), \tilde{\sigma}_{2}(r)(r^{*}), \\
  \tilde{\phi_{1}^{T}}(q^{*})(p), \tilde{\phi_{2}^{T}}(r^{*})(p)\}.
\end{multline}
For convenience, we denote the left-hand side and the right-hand side of the above inequality as $A(q)(r)(p)$ and $B(q)(r)(p)$, respectively.
On the other hand, by $\tilde{G}_{3}\subseteq_{\tilde{\phi}_{1}} \tilde{G}_{1}$ and $\tilde{G}_{3}\subseteq_{\tilde{\phi}_{2}} \tilde{G}_{2}$,
$\forall q\in[1, m], p\in[1, k]$, we have
\begin{multline}
  \max_{p^{*}\in[1, k]}\min\{\tilde{\phi}_{1}^{T}(q)(p^{*}), \tilde{\sigma}_{3}(p^{*})(p) \} \leq \\
  \max_{q^{*}\in[1, m]}\min\{ \tilde{\sigma}_{1}(q)(q^{*}), \tilde{\phi_{1}^{T}}(q^{*})(p)\},
\end{multline}
and  $\forall r\in[1, n], p\in[1, k]$,
\begin{multline}
  \max_{p^{*}\in[1, k]}\min\{\tilde{\phi}_{2}^{T}(r)(p^{*}), \tilde{\sigma}_{3}(p^{*})(p) \} \leq  \\
   \max_{r^{*}\in[1, m]}\min\{ \tilde{\sigma}_{2}(r)(r^{*}), \tilde{\phi_{2}^{T}}(r^{*})(p)\}.
\end{multline}
Suppose when $q^{*}=q^{0}$ and $r^{*}=r^{0}$, the right-hand side of Equations (32) and (33) get the maxima. Then
$\forall q\in[1, m], p\in[1, k]$, we get
\begin{multline}
  \max_{p^{*}\in[1, k]}\min\{\tilde{\phi}_{1}^{T}(q)(p^{*}), \tilde{\sigma}_{3}(p^{*})(p) \} \leq \\
  \min\{ \tilde{\sigma}_{1}(q)(q^{0}), \tilde{\phi_{1}^{T}}(q^{0})(p)\},
\end{multline}
and $\forall r\in[1, n], p\in[1, k]$,
\begin{multline}
  \max_{p^{*}\in[1, k]}\min\{\tilde{\phi}_{2}^{T}(r)(p^{*}), \tilde{\sigma}_{3}(p^{*})(p) \} \leq  \\
  \min\{\tilde{\sigma}_{2}(r)(r^{0}), \tilde{\phi_{2}^{T}}(r^{0})(p)\}.
\end{multline}
It is obvious that $\forall q\in[1, m], r\in[1, n], p\in[1, k]$, the following two equations hold:\par
\begin{gather}
A(q)(r)(p) \leq \max_{p^{*}\in[1, k]}\min\{\tilde{\phi}_{1}^{T}(q)(p^{*}), \tilde{\sigma}_{3}(p^{*})(p) \}, \\
A(q)(r)(p) \leq \max_{p^{*}\in[1, k]}\min\{\tilde{\phi}_{2}^{T}(r)(p^{*}), \tilde{\sigma}_{3}(p^{*})(p) \}.
\end{gather}
Based on Equations (34)-(37), we get
\begin{multline}
  A(q)(r)(p) \leq \min\{ \tilde{\sigma}_{1}(q)(q^{0}), \tilde{\sigma}_{2}(r)(r^{0}), \\
   \tilde{\phi_{1}^{T}}(q^{0})(p), \tilde{\phi_{1}^{T}}(r^{0})(p) \}.
\end{multline}
Moreover, it is obvious that
\begin{multline}
  B(q)(r)(p) \geq \min\{ \tilde{\sigma}_{1}(q)(q^{0}), \tilde{\sigma}_{2}(r)(r^{0}), \\
   \tilde{\phi_{1}^{T}}(q^{0})(p), \tilde{\phi_{1}^{T}}(r^{0})(p) \}.
\end{multline}
Then we get $A(p)(q)(r) \leq B(p)(q)(r)$. That is, we have shown Equation (31), which means $\tilde{\phi}^{T} \odot \tilde{\sigma}_{3} \leq \tilde{\sigma}_{1} \tilde{\otimes} \tilde{\sigma}_{2} \odot \tilde{\phi}^{T}$.
 \end{enumerate}
Therefore, we have completed the proof of Proposition 4.
%
\section{Proof of Proposition 5}
Suppose $\tilde{G}_{3}\subseteq_{\tilde{\phi}_{3}} \tilde{G}_{1} ||\tilde{G}_{2} $ and $|X_{1}| = m$, $|X_{2}| = n$, $|X_{3}| = k$. Let
 $\tilde{\phi}_{1}(p)(q) = \max_{r=1}^{m}\tilde{\phi}_{3}(p)((q-1)*n+r)$,  $\forall q\in[1, m], p\in[1, k]$.
  Here  $\tilde{\phi}(i)(j)$ denotes the $i$th row and $j$th column element of the matrix $\tilde{\phi}$.
  We prove $\tilde{G}_{3} \subseteq_{\tilde{\phi}_{1}} \tilde{G}_{1} $ as follows. \par
 \begin{enumerate}
 \item By $\tilde{G}_{3}\subseteq_{\tilde{\phi}_{3}} \tilde{G}_{1} ||\tilde{G}_{2}$, we have $\tilde{x}_{03}\leq (\tilde{x}_{01} \otimes \tilde{x}_{02})\odot \tilde{\phi}_{3}^{T}$. That is, $\forall~ p \in[1, k]$,
\begin{multline}
 \tilde{x}_{03}(p) \leq \max_{q^{*}\in[1, n]}^{r^{*}\in[1, n]}\min\{\min\{\tilde{x}_{01}(q^{*}), \tilde{x}_{02}(r^{*})\}, \\
 \tilde{\phi}_{3}^{T}((q^{*}-1)*n+r^{*})(p) \}. \\
 \end{multline}
 By the definition of $\tilde{\phi}_1$, we have
 \begin{equation}
 \tilde{\phi}_{3}^{T}((q^{*}-1)*n+r^{*})(p) \leq  \tilde{\phi}_{1}^{T}(q^{*})(p).
 \end{equation}
 Then we have
 \begin{equation}
  \tilde{x}_{03}(p) \leq \max_{q^{*}\in[i, n]}\min\{\tilde{x}_{01}(q^{*}), \tilde{\phi}_{1}^{T}(q^{*})(p) \}.
 \end{equation}
That is, $\tilde{x}_{03} \leq \tilde{x}_{01} \odot \tilde{\phi}_{1}^{T}$.
\item
By $\tilde{G}_{3}\subseteq_{\tilde{\phi}_{3}} \tilde{G}_{1} ||\tilde{G}_{2}$, we have $\tilde{x}_{m3}\odot \tilde{\phi}_{3} \leq \tilde{x}_{m1} \tilde{\otimes} \tilde{x}_{m2}$. That is, $\forall q\in[1, m], r\in[1, n]$,
\begin{multline}
  \max_{p^{*}\in[1, m]}\min\{\tilde{x}_{m3}(p^{*}), \tilde{\phi}_{3}(p^{*})((q-1)*n+r) \}\\
  \leq \min\{\tilde{x}_{m1}(q), \tilde{x}_{m2}(r)\}.
\end{multline}
Suppose $\tilde{\phi}_{1}(p^{*})(q)) = \tilde{\phi}_{3}(p^{*})((q-1)*n+r^{*}) $. Then we have
\begin{align}
&\max_{p^{*}\in[1, m]}\min\{\tilde{x}_{m3}(p^{*}), \tilde{\phi}_{1}(p^{*})(q)\} \nonumber \\
&= \max_{p^{*}\in[1, m]}\min\{\tilde{x}_{m3}(p^{*}), \tilde{\phi}_{3}(p^{*})((q-1)*n+r^{*})\} \nonumber \\
  &\leq \min\{\tilde{x}_{m1}(q), \tilde{x}_{m2}(r^{*})\} \leq \tilde{x}_{m1}(q).
\end{align}
That is, $\tilde{x}_{m3}\odot \tilde{\phi}_{1} \leq \tilde{x}_{m1}$.
\item By $\tilde{G}_{3}\subseteq_{\tilde{\phi}_{3}} \tilde{G}_{1} ||\tilde{G}_{2}$, we have $\tilde{\phi}_{3}^{T}\odot \tilde{\sigma}_{3} \leq (\tilde{\sigma}_{1} \tilde{\otimes} \tilde{\sigma}_{2}) \odot \tilde{\phi}_{3}^{T}$. That is,  $\forall q\in[1, m], r\in[1, n], p\in[1, k]$,
\begin{multline}
\max_{p^{*}\in[1, k]}\min\{ \tilde{\phi}_{3}^{T}((q-1)*n+r)(p^{*}), \tilde{\sigma}_{3}(p^{*})(p)\} \leq \\
\max_{q^{*}\in[1, m]}^{r^{*}\in[1, n]}\min\{  \tilde{\sigma}_{1}(q)(q^{*}), \tilde{\sigma}_{2}(r)(r^{*}),\\
 \tilde{\phi}_{3}^{T}((q^{*}-1)*n+r^{*})(p)\}.
\end{multline}
Suppose $\tilde{\phi}_{1}^{T}(q)(p^{*})) = \tilde{\phi}_{3}^{T}((q-1)*n+r^{0})(p^{*}) $. Then by the definition of $\tilde{\phi}_1$, we have
\begin{align*}
  & \max_{p^{*}\in[1, k]}\min\{\tilde{\phi}_{1}^{T}(q)(p^{*}), \tilde{\sigma}_{3}(p^{*})(p) \}\\
  &= \max_{p^{*}\in[1, k]}\min\{\tilde{\phi}_{3}^{T}((q-1)*n+r^{0})(p^{*}), \tilde{\sigma}_{3}(p^{*})(p)\} \\
  &\leq \max_{q^{*}\in[1, m]}^{r^{*}\in[1, n]}\min\{ \tilde{\sigma}_{1}(q)(q^{*}), \tilde{\sigma}_{2}(r^{0})(r^{*}), \\ &\tilde{\phi}_{3}^{T}((q^{*}-1)*n+r^{*})(p)\}\\
   &\leq \max_{q^{*}\in[1, m]}\min\{ \tilde{\sigma}_{1}(q)(q^{*}), \tilde{\phi}_{1}^{T}(q^{*})(p)\}.
\end{align*}
That is, $ \tilde{\phi}_{1}^{T} \odot \tilde{\sigma}_{3} \leq \tilde{\sigma}_{1} \odot \tilde{\phi}_{1}^{T}$.
\end{enumerate}

Therefore, the proof of Proposition 5 has been completed.

\section*{Acknowledgments}
The authors would like to thank the anonymous referees and area editor for their suggestions and comments to help us improve the quality of the paper.
This work is supported in part by the National
Natural Science Foundation (Nos. 61272058, 61073054), the Natural
Science Foundation of Guangdong Province of China (No.
10251027501000004),  the Specialized Research Fund for the Doctoral Program of Higher Education of China
(No. 20100171110042), and the project
of  SQIG at IT, funded by FCT PEst-OE/EEI/\linebreak[0]LA0008/2013.

\ifCLASSOPTIONcaptionsoff
  \newpage
\fi

\nocite{*}
\bibliographystyle{IEEEtran}
\bibliography{newref2}

\begin{IEEEbiography}[{\includegraphics[width=1in,height=1.25in,clip,keepaspectratio]{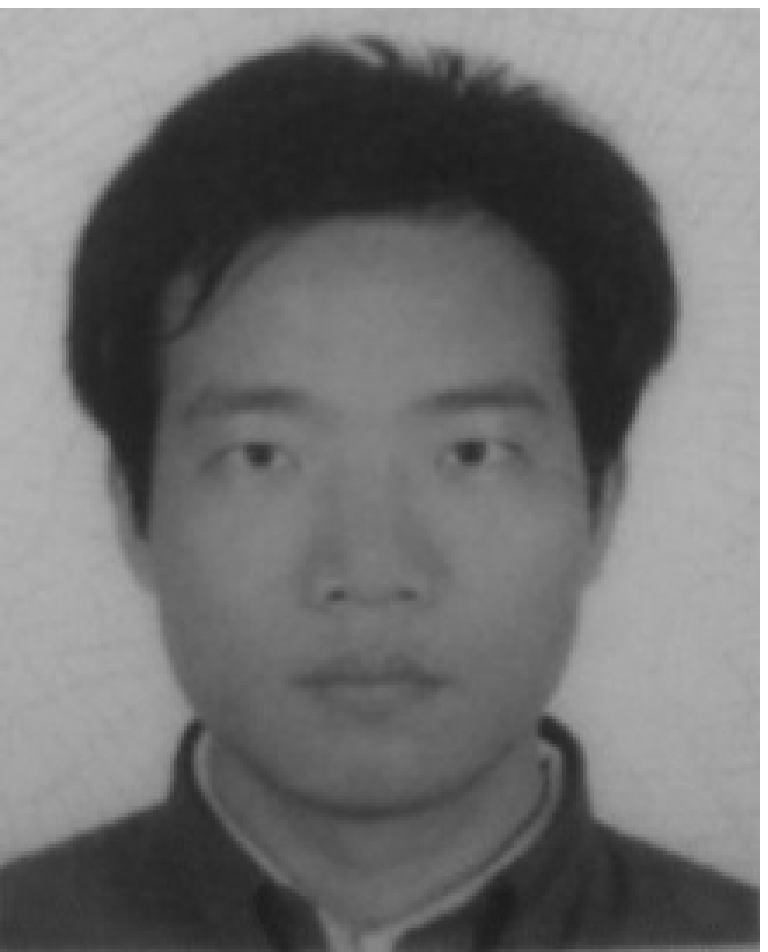}}]
{Weilin Deng}
received the B.S.  degree and the M.S. degree in computer science from the South China University of Technology, Guangzhou, China, in 2003 and 2008, respectively. Since 2011, he has been working toward the Ph.D. degree in the Department of Computer Science, Sun Yat-sen University, Guangzhou, China.
 \par
He is interested in fuzzy discrete event systems and control issues related.
\end{IEEEbiography}

\begin{IEEEbiography}[{\includegraphics[width=1.0in,height=1.25in,clip,keepaspectratio]{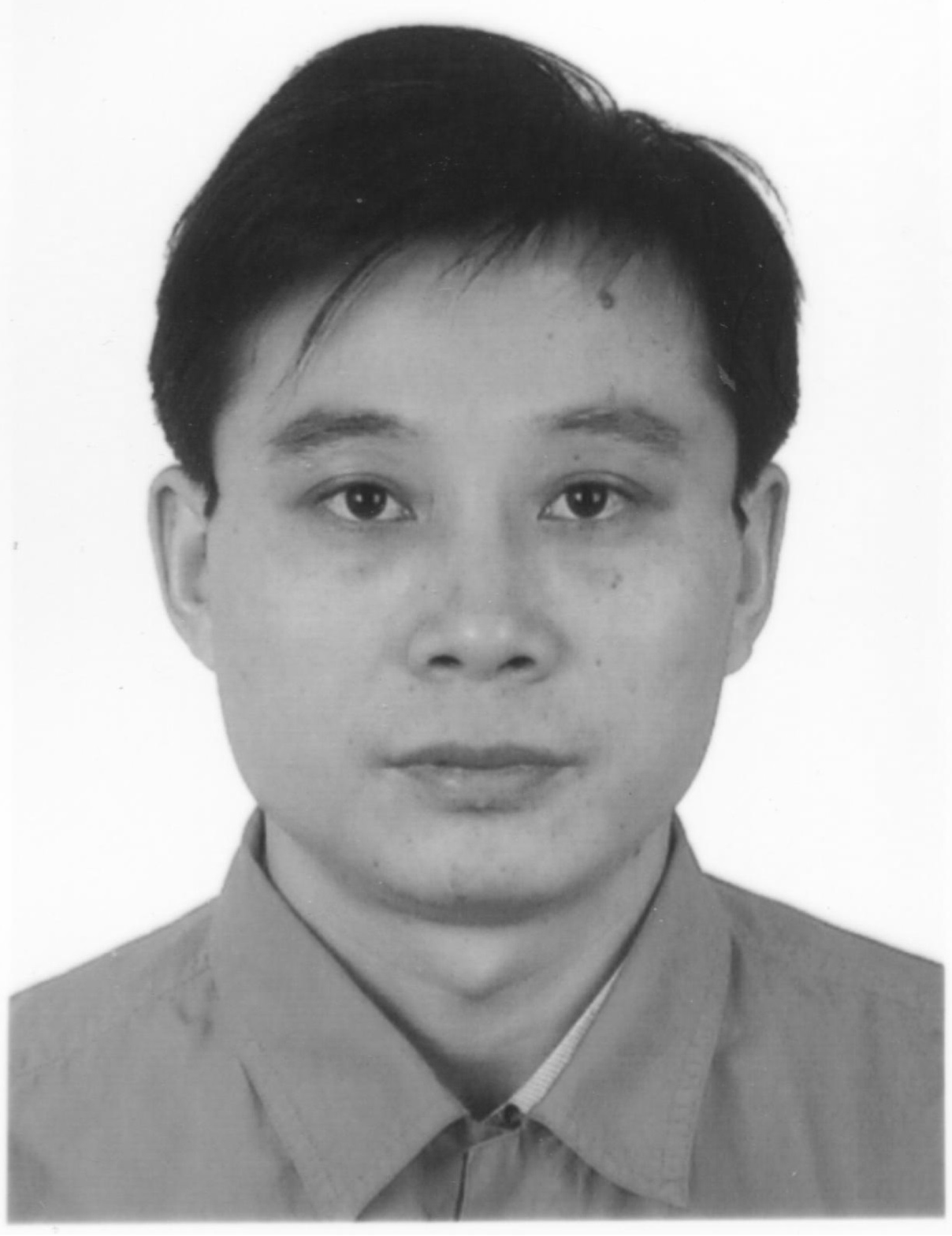}}]
{Daowen Qiu}

 received the M.S. degree in mathematics in 1993 from Jiangxi Normal University, Nachang, China, and then he received the Ph.D. degree in mathematics from Sun Yat-Sen University, Guangzhou, China, in 2000. He completed the postdoctoral research in computer science at Tsinghua University, Beijing, China, in 2002.

Since 2004, he has been a Professor of computer science at Sun Yat-Sen University. He is interested in models of nonclassical computation (including quantum, fuzzy and probabilistic computation) and quantum information theory. He has published over 100 peer-review papers in academic journals and international conferences, including: Information and Computation, Artificial Intelligence, Journal of Computer and System Sciences, Theoretical Computer Science, IEEE Transactions on Automatic Control, IEEE Transactions on SMC-Part B, IEEE Transactions on Fuzzy Systems, Physical Review A, Quantum Information and Computation, Journal of Physics A, Science in China.

\end{IEEEbiography}


\end{document}